\documentclass[preprint,12pt,authoryear]{elsarticle}

\usepackage{amsmath}
\usepackage{amssymb}

\usepackage{color}

\usepackage{subfig}

\usepackage{pdfpages}

\usepackage[ruled]{algorithm2e}

\begin{document}

\begin{frontmatter}

\title{Skew-Gaussian Random Fields}

\author[ntnu]{Kjartan ~Rimstad\corref{cor1}}
\ead{rimstad@gmail.com}

\author[ntnu]{Henning ~Omre}
\ead{omre@math.ntnu.no}

\address[ntnu]{Department of Mathematical Sciences, Norwegian University of Science and Technology, Trondheim, Norway}
\cortext[cor1]{Corresponding author: Department of Mathematical Sciences, NTNU, NO-7491 Trondheim, Norway. Tel.: +47  73 59 35 20; fax: +47 73 59 35 24}

% 
% \documentclass[a4paper,11pt]{elsarticle}
% \usepackage[utf8x]{inputenc}
% 
% \usepackage{amsmath}
% \usepackage{amsfonts}
% \usepackage{amssymb}
% 
% \usepackage[colorinlistoftodos]{todonotes}
% % \usepackage[disable,colorinlistoftodos]{todonotes}
% 
% \usepackage{url}
% \usepackage{natbib}
% % \usepackage[comma]{natbib}
% 
% \usepackage{geometry}
% 
% \usepackage{rotating}
% 
% \usepackage[lined,boxed,titlenotnumbered]{algorithm2e}
% 
% \usepackage{microtype}
% 
% \usepackage{subfig}
% 
% \usepackage{endfloat}
% 
% \usepackage{setspace}
% \doublespacing

% \title{Skew-Gaussian Random Fields \vspace{0.25cm}}
% \author{
%        Kjartan Rimstad \& Henning Omre \vspace{0.15cm} \\
% \small Department of Mathematical Sciences\\
% \small Norwegian University of Science and Technology\\
% \small Trondheim, Norway \vspace{0.35cm} \\
% Running head: Skew-Gaussian Random Fields 
% }
% \date{July 6, 2012}
% \date{}

% \begin{document}

% \maketitle

\begin{abstract}
Skewness is often present in a wide range of spatial prediction problems, and modeling it in the spatial context remains a challenging problem. In this study a skew-Gaussian random field is considered. The skew-Gaussian random field is constructed by using the multivariate closed skew-normal distribution, which is a generalization of the traditional normal distribution. We present a Metropolis-Hastings algorithm for simulating realizations efficiently from the random field, and an algorithm for estimating parameters by maximum likelihood with a Monte Carlo approximation of the likelihood. We demonstrate and evaluate the algorithms on synthetic cases. The skewness in the skew-Gaussian random field is found to be strongly influenced by the spatial correlation in the field, and the parameter estimators appear as consistent with increasing size of the random field. Moreover, we use the closed skew-normal distribution in a multivariate random field predictive setting on real seismic data from the Sleipner field 
in the North Sea. 
\end{abstract}

\vspace{0.5cm} 


\end{frontmatter}

\renewcommand{\Pr}{\operatorname{P}}
\newcommand{\E}{\operatorname{E}}
\newcommand{\Var}{\operatorname{Var}}
\newcommand{\Cov}{\operatorname{Cov}}
\newcommand{\Cor}{\operatorname{Cor}}
\newcommand{\tr}{\operatorname{tr}}

\newcommand{\argmax}{\operatornamewithlimits{argmax}}
\newcommand{\argmin}{\operatornamewithlimits{argmin}}

\clearpage

\section{Introduction}

Spatial prediction is an important element in many earth science and engineering applications, such as climate studies, geographical and geological sciences, petroleum engineering, mining, and environmental engineering. Usually, data are considered to be a realization from a random field, and focus is on predicting values in unobserved locations or regions. The random field is often assumed to be Gaussian, but histograms of the raw data are frequently skewed, multi-modal, and/or heavy tailed and hence do not appear as Gaussian. A common approach to deal with non-Gaussianity is to univariately transform the spatial field into a field with Gaussian marginals and then use Gaussian models, but the transformation is usually not known and the inverse transformation may be difficult to assess \citep[see e.g.][]{DeOliveira1997,Diggle2007}. Moreover it is difficult to create a joint transformation of the entire random field into a Gaussian random field. An alternative strategy is to consider a non-Gaussian random 
field model. 
We follow the latter approach in this study and define a random field which capture skewness. 

The univariate skew-normal distribution was introduced in \citet{Azzalini1985}, and the multivariate skew-normal distribution in \citet{Azzalini1996} and \citet{Azzalini1999}. Several authors have generalized the multivariate skew-normal distribution and many of these generalizations are summarized in \citet{Arellano-Valle2006}. The book \citet{Genton2004} provides a detailed overview over a variety of skewed probability distributions. We consider the multivariate closed skew-normal (CSN) distribution  introduced in \citet{Gonzalez-Farias2004,Gonzalez-Farias2004a}. The family of CSN distributions inherits many favorable properties from the multivariate normal distribution, it is for example closed under marginalization and conditioning. 

The skew-normal probability distributions have also previously been cast in a spatial context. In \citet{Kim2004} a skew-normal random field using the multivariate skew-normal distribution is defined, and an approach for spatial interpolation is presented. \citet{Allard2007} studies a random field constructed from the CSN distribution and outline a procedure for spatial prediction. Bayesian spatial prediction for CSN random fields is presented in \citet{Karimi2011}, and Bayesian spatial regression with CSN errors and missing observations is considered in \citet{Karimi2010}. \citet{Zhang2010} presents an approach based on the univariate skew-normal distribution, but the model lacks the closure properties which the CSN model exhibits. In the current study we follow \citet{Allard2007} and consider the random field constructed from the CSN distribution. This choice is made because the CSN distribution family has several favorable closure properties, which will be exploited for inference and 
prediction. We extend the study in \citet{Allard2007} by using a grid representation which defines an approximately stationary random field, using a slightly different parameterization, estimating model parameters by maximum likelihood, and using the model in a predictive setting with real seismic data from the North Sea. 

In order to evaluate the CSN density function we have to calculate high dimensional orthant probabilities of the normal distribution, which is an extremely computer demanding task. We present an algorithm inspired by the algorithms presented in \citet{Genz1992} and \citet{Genz2009} to calculate these orthant probabilities. Moreover, an efficient method for sampling from truncated normal distribution is presented. Lastly, we use a Monte Carlo based maximum likelihood estimation approach to assess the model parameters \citep{Geyer1992}. All the computer calculation are made on the laptop computer (Intel Core i7 CPU, 8 GB memory).

\section{Model}

The multivariate CSN distribution is defined in \citet{Gonzalez-Farias2004}, as a generalization of the multivariate normal distribution. Let $\mathbf U$ be multivariate normal distributed by using the notation:
\begin{align}
\mathbf U 
=
\left(
\begin{array}{c}
\mathbf U_1 \\
\mathbf U_2
\end{array}
\right)
\sim
N_{p+q} 
\left[
\left(
\begin{array}{c}
\boldsymbol  \mu_1 \\
\boldsymbol  \mu_2
\end{array}
\right)
,
\left(
\begin{array}{cc}
\mathbf  \Sigma_1 & \mathbf \Sigma_{12} \\
\mathbf  \Sigma_{21} & \mathbf \Sigma_2 \\
\end{array}
\right)
\right],
\end{align}
where $\mathbf U \in \mathbb{R}^{p+q}$, $\mathbf U_1, \boldsymbol \mu_1 \in \mathbb{R}^p$, $\mathbf U_2, \boldsymbol \mu_2 \in \mathbb{R}^q$, $\boldsymbol \Sigma_1 \in \mathbb{R}^{p\times p}$, $\boldsymbol \Sigma_2 \in \mathbb{R}^{q\times q}$, $\boldsymbol \Sigma_{12} = \boldsymbol \Sigma_{21}^T \in \mathbb{R}^{p\times q}$, and $N_{n}(\boldsymbol \mu, \boldsymbol \Sigma)$ denotes the $n$-dimensional multivariate normal distribution with expectation vector $\boldsymbol \mu$ and covariance matrix $\boldsymbol \Sigma$. Then  $\mathbf X = [\mathbf U_1 \mid \mathbf  U_2 \leq \mathbf 0 ]$ is defined to be CSN distributed, denoted $\mathrm{CSN}_{p,q}(\boldsymbol \mu_1, \boldsymbol \Sigma_1, -\boldsymbol \Sigma_{21} \boldsymbol \Sigma_1^{-1}, \boldsymbol \mu_2, \boldsymbol \Sigma_2 - \boldsymbol \Sigma_{21} \boldsymbol \Sigma_1^{-1} \boldsymbol \Sigma_{12})$. Here the notation $\mathbf U_2 \leq \mathbf 0$ corresponds to all elements in $\mathbf U_2$ being jointly negative. The corresponding probability density 
function (pdf) of the CSN distribution is
\begin{align}
&f_{p,q} (\mathbf x;\boldsymbol \mu_1, \boldsymbol \mu_2, \boldsymbol \Sigma_1, \boldsymbol \Sigma_2, \boldsymbol \Sigma_{12}) \notag \\
= \; 
& \phi_p(\mathbf x;\boldsymbol \mu_1, \boldsymbol \Sigma_1)  \;
\frac{
\Phi_q(\mathbf 0; \boldsymbol \mu_2 + \boldsymbol \Sigma_{21} \boldsymbol \Sigma_1^{-1} (\mathbf x - \boldsymbol \mu_1), \boldsymbol \Sigma_2 - \boldsymbol \Sigma_{21} \boldsymbol \Sigma_1^{-1} \boldsymbol \Sigma_{12})
}{
\Phi_q(\mathbf 0; \boldsymbol \mu_2, \boldsymbol \Sigma_2)
}\label{eqn:csn_density},
\end{align}
where $\phi_\cdot(\cdot;\cdot,\cdot)$ is the normal pdf and $\Phi_\cdot (\cdot; \cdot, \cdot)$ is the normal cumulative distribution function (cdf).  For $\boldsymbol \Sigma_{12}$ being a matrix of zeros, the CSN projects into the multivariate normal distribution. For $q= 1$ and $\boldsymbol \mu_2=\mathbf 0$ the CSN is identical to the multivariate skew-normal distribution as defined in \citet{Azzalini1996}. The class of CSN distributions inherits important properties from the multivariate normal distribution, such as being closed under marginalization, conditioning, and linear transformations \citep{Gonzalez-Farias2004}.

We work in a spatial random field context and we let $\left\lbrace Z(\mathbf s): \mathbf s \in \mathcal D  \right\rbrace$ be a real-valued random field, where $\mathcal D$ is a finite domain in $\mathbb R^d$ and $\mathbf s \in \mathbb R^d$ is a generic location in $\mathcal D$. The random field $\left\lbrace Z(\mathbf s): \mathbf s \in \mathcal D \right\rbrace$ is a Gaussian random field if for all configurations of locations $(\mathbf s_1,\ldots, \mathbf s_n)$ and all $n>0$ the pdf of $\mathbf Z = [Z(\mathbf s_1), \ldots, Z(\mathbf s_n)]^T$ is a multivariate normal distribution.

Consider the bivariate Gaussian random field 
\begin{align}
\left\lbrace 
\mathbf U(\mathbf s)
=
\left(
\begin{array}{c}
U_1(\mathbf s) \\
U_2(\mathbf s)
\end{array}
\right): \ \ 
\mathbf s \in \mathcal D
\right\rbrace. \label{eqn:cont_joint}
\end{align}
Then the random field defined by
\begin{align}
\left\lbrace Y(\mathbf s) =  \left[ U_1(\mathbf s) \mid U_2(\mathbf s') \leq 0 : \; \mathbf s' \in \mathcal D\right] : \mathbf s \in \mathcal D \right\rbrace \label{eqn:cont_skew_field},
\end{align}
is a skewed Gaussian random field. One particular case occurs when $U_1(\mathbf s)$ and $U_2(\mathbf s)$ are independent, then $Y(\mathbf s)$ is a Gaussian random field. Moreover, for the extreme case with full dependence we have that $Y(\mathbf s)$ is a truncated Gaussian random field. When the bivariate $\mathbf U(\mathbf s)$ is a stationary Gaussian random field, then $Y(\mathbf s)$ is also a stationary random field, except for border effects.

The random field defined in Expression \ref{eqn:cont_skew_field} is difficult to handle in practice; therefore, we consider a CSN random field as defined in \citet{Allard2007}. The random field is defined by first specifying a discretization $(\mathbf s'_1, \ldots,\mathbf s'_q)$ with finite and fixed $q$, and $\mathbf U_2 = \left[ U_2(\mathbf s'_1), \ldots, U_2(\mathbf s'_q) \right]$. Then we define the associated CSN random field as 
\begin{align}
\left\lbrace X(\mathbf s) =  \left[ U_1(\mathbf s) \mid \mathbf U_2 \leq \mathbf 0 \right] : \mathbf s \in \mathcal D \right\rbrace, \label{eqn:csn_field}
\end{align}
if for all configurations of locations $(\mathbf s_1,\ldots, \mathbf s_p)$ and all $p>0$ the pdf of $\mathbf X = [X(\mathbf s_1), \ldots, X(\mathbf s_p)]^T$ is CSN distributed. Or equivalently, if the Gaussian random field $U_1(\mathbf s)$ and $\mathbf U_2$ are jointly Gaussian, then $\left\lbrace X(\mathbf s) = \right.$ $ \left[ U_1(\mathbf s) \mid \mathbf U_2 \leq \mathbf 0 \right] :$ $\left.\mathbf s \in \mathcal D \right\rbrace$ is a CSN random field.

Note that even when $U(\mathbf s)$ in Expression \ref{eqn:cont_joint} is a stationary Gaussian random field, the CSN random field as defined in Expression \ref{eqn:csn_field} is not stationary in the traditional sense. The non-stationarity is a consequence of the locations $(\mathbf s'_1, \ldots,\mathbf s'_q)$ being finite and fixed. For example when $\mathbf s$ is far from all $(\mathbf s'_1, \ldots,\mathbf s'_q)$ and hence $U(\mathbf s)$ and $\mathbf U_2$ are weakly correlated, then the marginal distribution of $U(\mathbf s)$ will be approximately Gaussian. 

The random field has some stationary properties, however, for example for $(\mathbf s'_1, \ldots,\mathbf s'_q)$ being a regular grid discretization over $\mathcal D$ the random field $ X(\mathbf s)$ is approximately stationary in the discretization locations $(\mathbf s'_1, \ldots, \mathbf s'_q)$ sufficiently far away from the border. In addition the CSN random field is approximately stationary when the discretization $(\mathbf s'_1, \ldots, \mathbf s'_q)$ in some sense is dense compared to the correlation range of the random field. The CSN random field can also be interpreted in a Bayesian setting, with a truncated random field $\left[ \mathbf U_2 \mid \mathbf U_2 \leq \mathbf 0 \right]$ seen as a latent random field \citep{Liseo2003}.

The skew-Gaussian random field introduced in \citet{Kim2004} is a special case of the CSN random field defined above with $q=1$. By choosing $q=1$, we obtain the CSN random field
\begin{align}
\left\lbrace  X(\mathbf s)  =  U_1(\mathbf s)  + \Cov(U_1(\mathbf s),U_2) \; \Sigma_{2}^{-1} ([U_2\mid U_2 \leq 0] - \mu_2): \quad \mathbf s \in \mathcal D \right\rbrace ,
\end{align}
where $U_1(\mathbf s)$ is a Gaussian random field and $[U_2\mid U_2 \leq 0]$ is a zero-truncated Gaussian random variable. If we assume that $\Cov(U_1(\mathbf s),U_2)$ is independent of $\mathbf s$ is it easy to see that this random field behaves like a Gaussian random field with a skewed mean \citep{Allard2007,Zhang2010} and the skewness will only be identified through repeated sampling from the random field. This effect is also observed for multivariate t-fields, where each realization behaves like a Gaussian random field, and the heavy tails will only be identifiable through repeated sampling from the random field \citep[see e.g.][]{Roislien2006}. Another consequence of each sample behaving like a Gaussian random field is that the skew-Gaussian random field model in \citet{Kim2004} cannot be identified by a single realization \citep{Genton2012}.

An additional problem with the skew-Gaussian random field introduced in \citet{Kim2004} is that $U_1(\mathbf s)$ contains the spatial variability and the scalar $[U_2\mid U_2 \leq 0]$ the skewness. If we fix the correlation structure of $U_1(\mathbf s)$ and assume that $U_1(\mathbf s)$ has short correlation length relative to the size of $\mathcal D$, the correlation between $U_1(\mathbf s)$ and $[U_2\mid U_2 \leq 0]$ has to be small. Thus the skewness in $X(\mathbf s)$ will be small. This is also discussed in \citet{Azzalini1996} where they consider the case when $p=2$ and $q=1$. Evidently similar problems will appear for the CSN random field when $p$ is large compared to $q$. In \citet{Karimi2011,Karimi2010} $q=2$ is used in the prediction part; thus each realization of the random field can not contain a high degree of skewness. 

In the current study we consider a two-dimensional approximately stationary CSN random field defined on a regular grid $\mathcal{L_D}$ over $\mathcal D \in \mathbb{R}^2$, which has the distribution in Expression \ref{eqn:csn_density}. We choose to use a parsimonious model with few parameters such that we are able to estimate the parameters from one realization of the random field. The model should, however, be sufficiently flexible such that it is able to describe different levels of skewness. We use $q = p$, where $(\mathbf s_1, \ldots,\mathbf s_p) = (\mathbf s'_1, \ldots,\mathbf s'_p)$ and $\mathbf s_i, \mathbf s'_i \in \mathcal{L_D}, \; i = 1,\ldots,p$, with location parameters $\boldsymbol \mu_1 = \mu \mathbf 1$, and $\boldsymbol \mu_2 = \nu \mathbf 1$, where $\mathbf 1 \in \mathbb R^p$ is a vector of ones. The covariance structure is defined to be on the form
\begin{align}
\boldsymbol \Sigma 
&=
\left[
\begin{array}{cc}
\sigma^2 \mathbf C 		& -\gamma \sigma \mathbf C \\
-\gamma \sigma \mathbf C	& (1-\gamma^2) \mathbf I_p + \gamma^2  \mathbf C
\end{array}
\right], \label{eqn:covariance}
\end{align}
where $\sigma^2>0$ is a scale parameter, $|\gamma| < 1$ is a skewness parameter, $\mathbf I_p$ is a $p$-dimensional identity matrix, and $\mathbf C$ is a stationary correlation matrix with an exponential correlation function $\rho(\boldsymbol \tau) = \mathrm{exp} \left\lbrace -\tau_1/d_h -\tau_2/d_v \right\rbrace$ where $\boldsymbol \tau = (\tau_1,\tau_2)$ is the distance between two spatial locations, and $d_h$ and $d_v$ are horizontal and vertical range parameters, respectively. The parameterization structure of $\boldsymbol \Sigma$ and the restrictions on the parameters ensure positive semidefiniteness of $\boldsymbol \Sigma$. The new parameterization of Expression \ref{eqn:csn_density} becomes
\begin{align}
f_{p,p} (\mathbf x; \mu, \nu, \sigma^2, \gamma, d_h, d_v) 
&=
\phi_p(\mathbf x; \mu \mathbf 1, \sigma^2 \mathbf C)  \;
\frac{
\prod_{i=1}^p \Phi_1(0;\nu - \frac{\gamma}{\sigma} ( x_i - \mu ), 1-\gamma^2 )
}{
\Phi_p(\mathbf 0;\nu \mathbf 1, (1-\gamma^2) \mathbf I_p + \gamma^2 \mathbf C)
}
, \label{eqn:param_density}
\end{align}
which is equivalent to $CSN_{p,p}(\mu \mathbf 1, \sigma^2 \mathbf C, \frac{\gamma}{\sigma} \mathbf I_p, \nu \mathbf 1, (1-\gamma^2) \mathbf I_p)$. The factorization in the nominator is a consequence of the identity matrix term in Expression \ref{eqn:covariance}. The parameters in the model are $\mu$, $\nu$, $\sigma^2$, $\gamma$, $d_h$, and $d_v$, and the constraints $\sigma^2,d_h,d_v > 0$ and $|\gamma| < 1$ ensure that $\boldsymbol \Sigma$ is positive semidefinite and hence a valid covariance matrix. Note that there is only one multivariate normal cdf $\Phi_p(\cdot;\cdot,\cdot)$ in Expression  \ref{eqn:param_density}. The marginal distribution for $x_j$ is 
\begin{align}
& \; f_{1,p} (x_j ; \mu, \nu, \sigma^2, \gamma, d_h, d_v) \notag \\
= & \;
\phi_1(x_j; \mu, \sigma^2) 
\Phi_1 (0;\nu - \frac{\gamma}{\sigma} (\mathbf x_j - \mu ), 1-\gamma^2 ) \notag \\
\times &  \; 
\frac{
\int 
\phi_{p-1}(\mathbf x_{-j} \mid x_j; \mu \mathbf 1, \sigma^2 \ \mathbf C)  \;
\prod_{i=1,i \neq j}^p \Phi_1(0;\nu - \frac{\gamma}{\sigma} ( x_i - \mu ), 1-\gamma^2 )
\mathrm d \mathbf x_{-j}
}{
\Phi_p(\mathbf 0;\nu \mathbf 1, (1-\gamma^2) \mathbf I_p + \gamma^2 \ \mathbf C)
}, \label{eqn:marginal}
\end{align}
where $\mathbf x_{-j} = [x_1,\ldots,x_{j-1},x_{j+1},\ldots,x_p]^T$ and $\phi_{p-1}(\mathbf x_{-j} \mid x_j; \mu \mathbf 1, \sigma^2 \ \mathbf C)$ is the conditional distribution of $\mathbf x_{-j}$ given $x_j$. This marginal distribution is not equal to the univariate skew-normal distribution introduced in \citet{Azzalini1985}. The marginal pdf in Expression \ref{eqn:marginal} is dependent on the complete grid design, hence all parameter inference and predictions must be made with reference to one common design. It is not so for Gaussian random fields where the marginal pdfs are independent of the spatial correlation function. These important features of CSN random fields are not discussed in \citet{Allard2007}. We will later in this study see that coupling between $x_j$ and $\mathbf x_{-j}$ will reduce the maximum skewness we can obtain in the marginal distribution compared to the corresponding skewness in the univariate skew-normal distribution introduced in \citet{Azzalini1985}. The maximum skewness is 
dependent on the spatial coupling of the grid nodes, and increased coupling decreases the maximum skewness. Increased coupling can be caused by either a denser grid or stronger spatial correlation, or both.  Moreover the random field is stationary except for border effects. The parameterization in Expression \ref{eqn:covariance} is similar, but not identical, to the parameterization in \citet{Allard2007} where the full covariance function is of the form
\begin{align}
\left[
\begin{array}{cc}
\sigma^2 \mathbf C 		& -\gamma \sigma^2 \mathbf C \\
-\gamma \sigma^2 \mathbf C	& (1+\gamma^2)\sigma^2 \mathbf C
\end{array}
\right].
\end{align}
Note that for this parameterization there are no restrictions on $\gamma$, but we will later in this study argue for our parameterization usually being able to capture a higher degree of skewness than the model in \citet{Allard2007}.

\subsection{Simulation study}%

In this section we explore the properties of the CSN random field defined above for six sets of parameter values, one base case, case 1, and five deviating cases. We are particularly concerned about the ability to represent skewness in the marginal pdf and its dependence on spatial coupling. The parameter values of the six cases are summarized in Table \ref{tbl:synthetic_param}, and we present $(50 \times 50)$ grid random fields for the six parameter cases. 

\begin{table}
\begin{center}
\begin{tabular}{c|cccccc}
Case & $\mu$ & $\nu$ & $\sigma^2$ & $\gamma$ & $d_h$ & $d_v$\\ \hline
1 & 0 & 0 & 1 & 0.975 & 3 & 3  \\
2 & 0 & 0 & 1 & 0.975 & 0 & 0  \\
3 & 0 & 0 & 1 & 0.975 & 5 & 5  \\
4 & 0 & 0 & 1 & 0.995 & 3 & 3  \\
5 & 0 & 2 & 1 & 0.975 & 3 & 3  \\
6 & 0 & 0 & 1 & 0.975 & 5 & 0 
\end{tabular}
\end{center}
\caption{Simulation study of CSN random field model. Model parameter values for six cases.}
\label{tbl:synthetic_param}
\end{table}

A Metropolis Hastings (MH) algorithm is used to sample from the distributions. The algorithm is summarized in Appendix A, and the algorithm uses the importance sampler from \cite{Genz1992} as a block proposal distribution in the update steps. The size of the blocks used in the MH-algorithm is normally about $100$ which gives an acceptance rate of about $0.23$ for most of the parameter cases we present. It takes a couple of minutes on a laptop computer to sample one realization of the CSN random field with the MH-algorithm implemented in C. The burn-in and mixing appear as satisfactory and is not displayed. 

\begin{figure}
  \begin{center}
    \includegraphics[height=0.9\textheight]{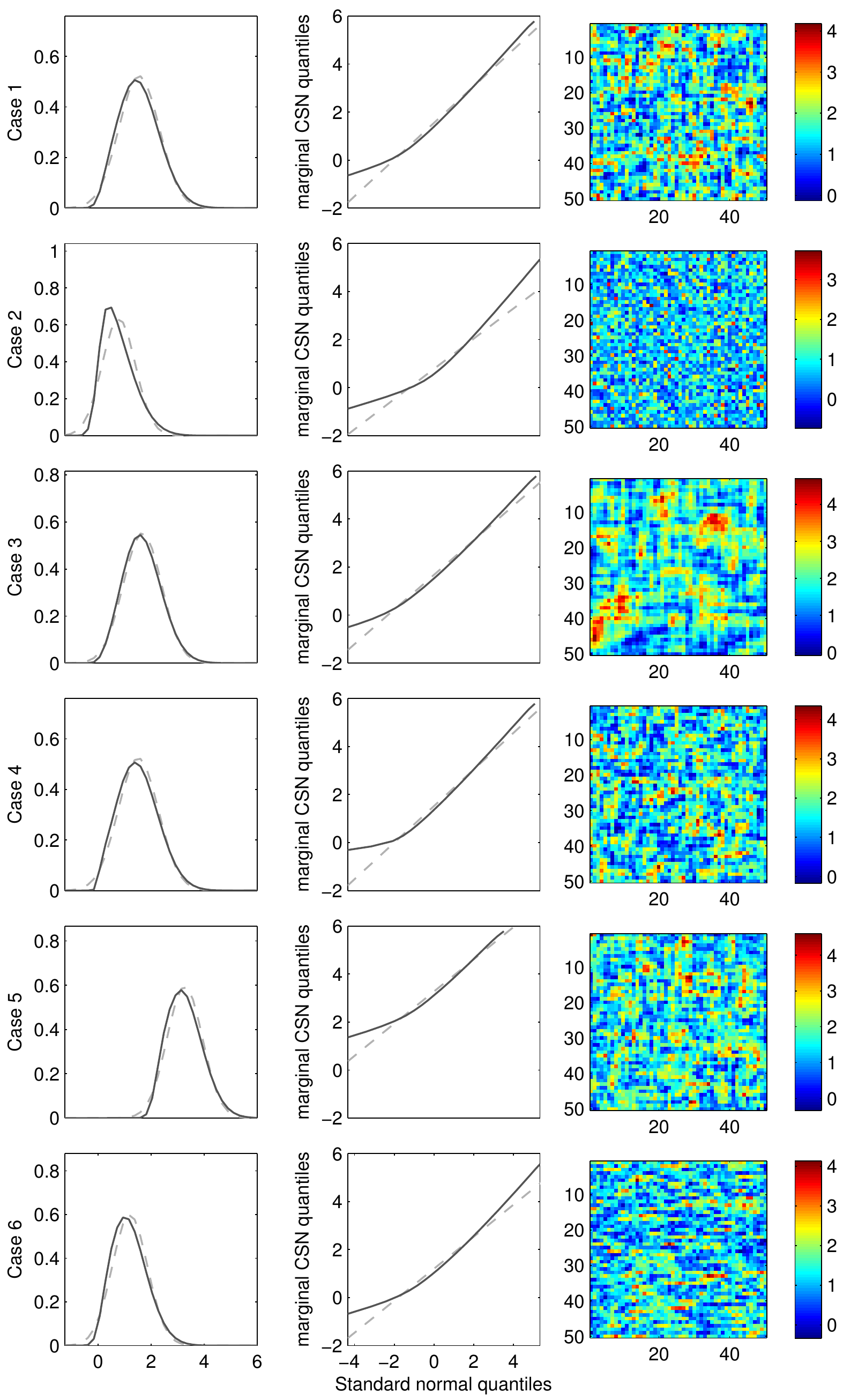}
    \caption{Characteristics of six cases of CSN random fields, see Table 1.  First column: black solid is the marginal distribution of CSN random field and dashed is the standard normal distribution. Second column: quantile-quantile plot of the marginal CSN random field versus theoretical quantiles from the normal distribution. Third column: one realization from the CSN random field.}
    \label{fig:synthetic_cases}
  \end{center}
\end{figure}

Figure \ref{fig:synthetic_cases} displays the results from the six cases. The marginal distributions in the center location of the random field are presented in the first column. Normal distributions, with the two first moments identical to the CSN marginal distribution, are also displayed. The second column displays quantile-quantile plots of the marginal CSN distribution versus the normal distribution. One arbitrary realization from the CSN random field is presented in the last column. 

The first row in Figure \ref{fig:synthetic_cases} displays the base case of a CSN random field with isotropic spatial correlation and reasonably strong correlation with the hidden truncated random field. Note that some skewness in the marginal distribution is visible, but the skewness is not very evident. In the second row a CSN random field without spatial correlation, i.e. a white noise random field, and otherwise identical parameters is displayed. We observe that the reduction in spatial correlation increases the skewness in the random field. 

The integral in Expression \ref{eqn:marginal} provides the spatial coupling effect since increased correlation of the variables make the mode of the truncated pdf move away from the truncation border and hence appear more normal like. Support S1 contains an illustrative example of the effect. This effect is a similar effect to the correlation versus skewness effect discussed in \citet{Azzalini1996}. The model in \citet{Allard2007} will usually have higher correlation 
in the truncated field than our model, due to the lack of the $(1-\gamma^2)\mathbf I_p$ term. Thus, the skewness in their model will be reduced compared to the skewness in our model.

In case 3, the third row of Figure \ref{fig:synthetic_cases}, the spatial correlation parameters are increased and otherwise identical parameters as case 1, which reduce the skewness even further. In the forth row the correlation parameter $\gamma$ between $\mathbf U_1$ and $\mathbf U_2$ is increased. Note that a value close to unity represents a random field that is close to a truncated Gaussian random field. The increase in $\gamma$ causes somewhat higher skewness in the marginal distribution. The fifth row presents results from a case where the truncation of the latent random field appears further out in the tail. This change of conditioning causes only minor changes in the marginal distribution compared to the base case. The last row in Figure \ref{fig:synthetic_cases} presents the case with spatial anisotropic correlation, with correlation only in the horizontal direction, and otherwise identical parameters to case 1. More skewness in the marginal distribution is obtained compared to case 1 where 
correlation 
is present in both directions. This spatial anisotropic correlation structure is similar to the one estimated for the seismic data case studied later in this paper. 

The study shows that it is difficult to obtain a high degree of skewness in the marginal distribution in CSN random fields due to the spatial coupling effect. This lack of skewness is unfortunate and reduce the relevance of the family of CSN random fields. Moreover, spatial coupling complicates model parameter estimation by a maximum likelihood criterion.

\section{Parameter estimation}
Parameter estimation for CSN random fields poses challenging numerical problems, since the probability density function and moments \citep[see][]{Allard2007} are functions of multivariate normal cdfs. In \citet{Allard2007} a method of moment estimation is discussed while a weighted method of moments estimation approach is used in \citet{Flecher2009}. The methods of moment estimators are particularly computational demanding due to the frequent appearance of the multivariate normal cdfs in the moments. In the current study we use a maximum likelihood estimator. The multivariate normal cdf also appears in the normalizing constant in the likelihood, but the total number of evaluations will usually be smaller than in a method of moments estimation procedure. From the previous section we learned that the differences between case 1 and case 5 are small. The only difference between the two cases is different values of $\nu$; therefore, we choose to fix $\nu=0$. In this section we only consider isotropic random 
fields and let $d = d_h = d_v$. The log-likelihood is then
\begin{align}
&l(\mu, \sigma^2, \gamma, d; \mathbf x) = \log L(\mu, \sigma^2, \gamma, d; \mathbf x) \notag \\
=&
\log \phi_p(\mathbf x; \mu \mathbf 1, \sigma^2 \ \mathbf C)  \;
+ \sum_{i=1}^q \log \Phi_1(0; - \frac{\gamma}{\sigma}  ( x_i - \mu ), 1-\gamma^2 ) \notag \\
&- \log \Phi_q(\mathbf 0; \mathbf 0, (1-\gamma^2) \mathbf I_p + \gamma^2 \mathbf C)
, \label{eqn:param_likelihood}
\end{align}
with $\mathbf C$ being the stationary correlation matrix as defined previously with $d = d_h = d_v$. We only need to compute the multivariate normal cdf once for each likelihood evaluation, while the optimization requires sequential likelihood computations. Note that for our model the challenging last term in Expression \ref{eqn:param_likelihood} only depends on the parameters $\gamma$ and $d$, not the parameters $\mu$ and $\sigma^2$, which simplifies the computations somewhat.

The model can be considered as a missing data model \citep{Little1987}, $\mathbf U_1$ is observed if $\mathbf U_2\leq0$, and the expectation–maximization \citep[EM,][]{Dempster1977} algorithm could be used to obtain the maximum likelihood estimate. Usually the EM-algorithm has slow convergence, and we cannot calculate the expectation step analytically; thus a Monte Carlo EM-algorithm \citep{Wei1990} has to be applied, which is a computational burden. Direct maximization of the likelihood requires estimates of multivariate normal cdfs that can be done by Monte Carlo methods, which is the approach chosen in this paper. Our experience is that this approach will normally converge faster than the alternatives discussed above.

Importance sampling provides a simple method for estimating the orthant probability $\Phi_q(\mathbf 0; \cdot, \cdot)$ in Expression \ref{eqn:param_likelihood}, and Monte Carlo optimization can be used to maximize the likelihood \citep{Geyer1996}. In our study we approximate these orthant probabilities by using the importance sampling method described in \citet{Genz1992} and \citet{Genz2009}, and the importance sampling algorithm is summarized in Appendix B. By using the same set of uniform random variables for each likelihood evaluation we ensure that the approximated likelihood is smooth; thus we are able to use standard optimization routines. The Monte Carlo errors in the Monte Carlo maximum likelihood estimates are evaluated by doing an ensemble of optimizations with different Monte Carlo likelihood approximations by using different set of random numbers. 

As discussed in \citet{Azzalini1985} the information matrix for the original parameterization of the univariate skew normal distribution becomes singular as the skewness parameter goes to zero, but this singularity problem can be solved by reparameterizing the model \citep{Azzalini1985,Azzalini1999}. The singularity in the information matrix becomes a problem in numerical optimization with Newton methods. We used the original parameterization of the CSN random field, but we used some steps with a Nelder-Mead simplex method before we used a interior-reflective Newton method in MATLAB to maximize the likelihood. By following this approach we did not notice any abnormalities in the optimization procedure.

The likelihood function is, however, generally not a convex function of the parameters. Thus to find the global optimum we start the optimization at multiple starting points and choose the values of the parameters that give the largest value of the likelihood function. In our simulation study we did not encounter any problem in identifying the dominant mode.

\subsection{Empirical study}

In this section we estimate parameters from realizations of CSN random field from the base case in the previous section, i.e. $\mu = 0$, $\sigma^2 = 1$, $\gamma = 0.975$, and $d=3$.  We consider the parameter estimates as a function of the dimension of the random field $p$ and the number of Monte Carlo samples $N$ used to assess the orthant probability. We want to evaluate the size of the random field needed and the computational demands required to get proper estimates. The computing time for $p=30^2$ and $N=1000$ is typically one minute on the laptop computer for estimating one set of parameters. 

\begin{figure}
  \begin{center}
    \includegraphics[width=1\textwidth]{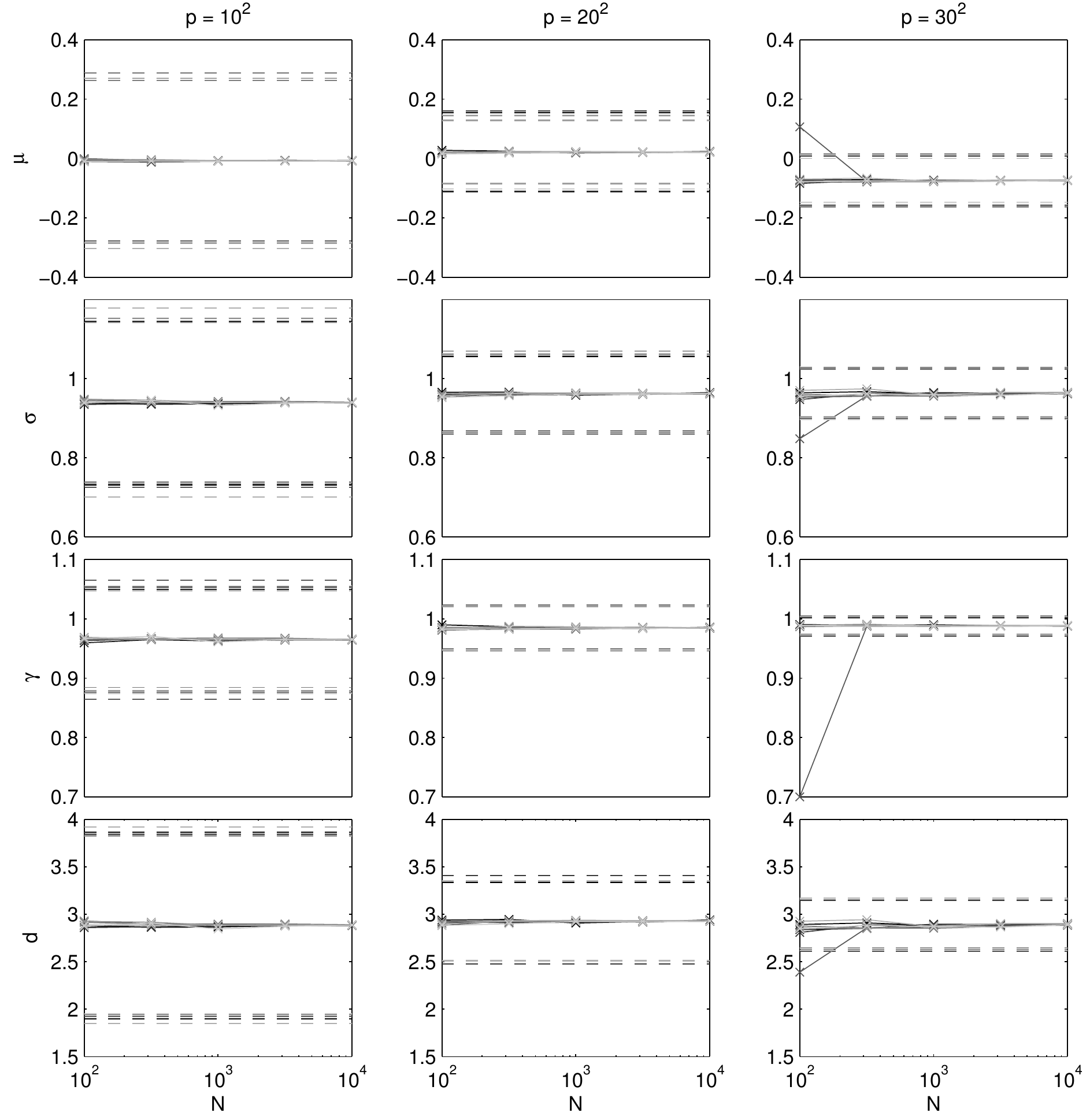}
    \caption{Empirical study of model parameter estimates by approximate maximum likelihood. Effect of Monte Carlo error for grid sizes  $p = $ $10^2$, $20^2$, $30^2$.}
    \label{fig:synthetic_mc_error}
  \end{center}
\end{figure}

Figure \ref{fig:synthetic_mc_error} displays the parameter estimates as a function of the number of Monte Carlo points $N$. The maximum likelihood estimates are plotted for eight different Monte Carlo approximations of the likelihood, i.e. we use eight sets of independent random numbers in the approximation of the likelihood. We use the same realization of the random field for all eight Monte Carlo likelihood approximations. The variations among these eight estimates illustrate the Monte Carlo likelihood approximation error in the parameter estimates. The dashed lines are $90\%$ prediction interval computed, where the variances are the diagonal elements of the inverse Hessian of the log likelihood function evaluated at its maximum. Figure \ref{fig:synthetic_mc_error} illustrates that the Monte Carlo error is significantly smaller than the prediction intervals for $N>100$, and that the Monte Carlo error increases with higher dimension $p$, while the inverse Hessian variance error decreases with higher 
dimension 
$p$. Hence we need higher value of $N$ for increasing $p$, which is expected since we need to compute orthant probabilities of higher dimensions.

\begin{figure}
  \begin{center}
    \includegraphics[width=\textwidth]{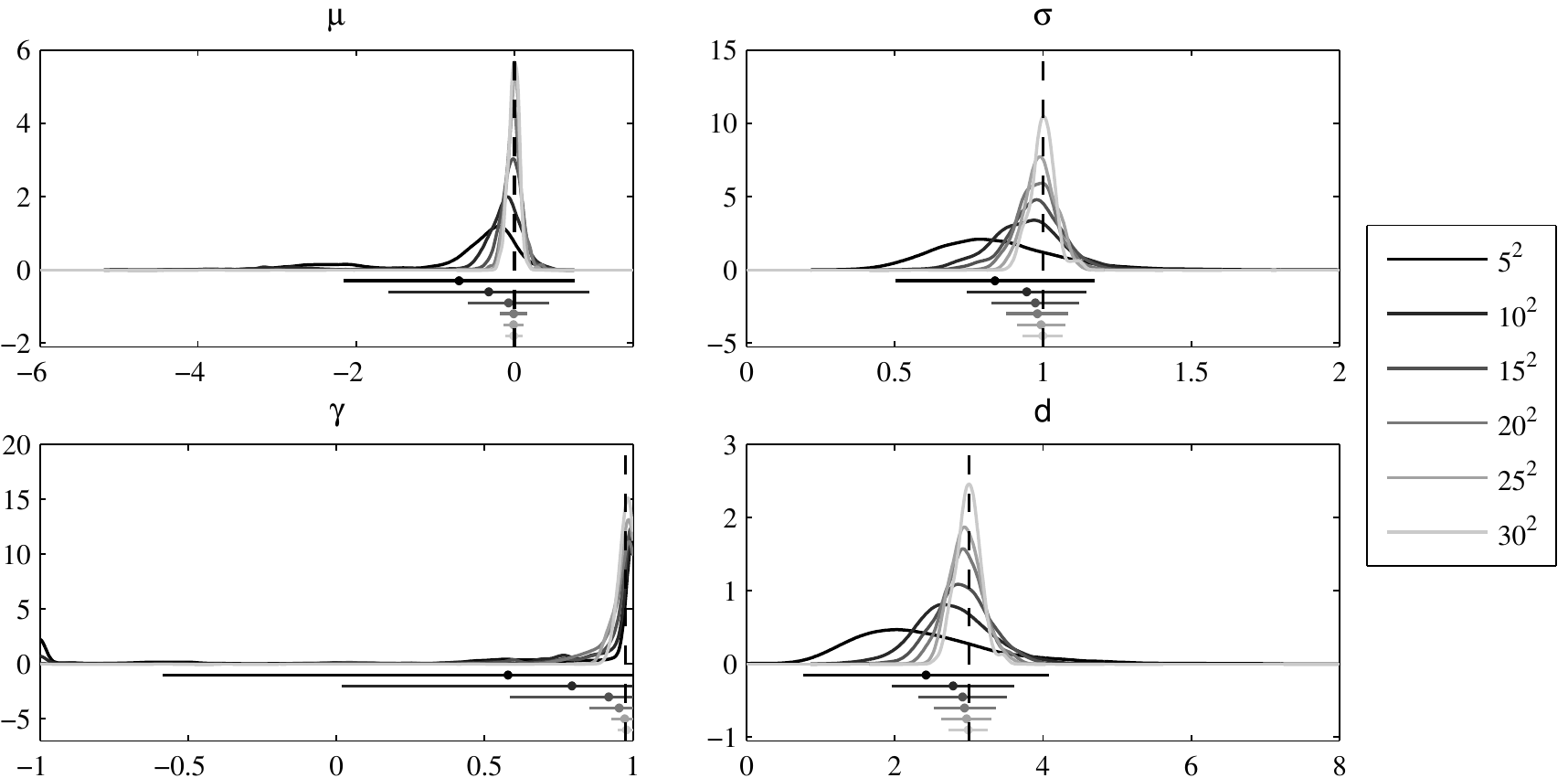}
    \caption{Empirical study of model parameter estimates by approximate maximum likelihood. Effect of increasing size of the observed field $p = 5^2$ to $30^2$. Displays of smoothed histogram based on $N = 1000$ Monte Carlo samples and $5000$ simulations. Also displayed, mean together with a 95\% prediction interval, and true values marked with a vertical dashed line.}
    \label{fig:synthetic_multiple}
  \end{center}
\end{figure}

Figure \ref{fig:synthetic_multiple} contains the distribution of the maximum likelihood estimates randomized over $5000$ realizations from the base case random field. The size of the random fields varies from $p=5^2$ to $p=30^2$. The number of Monte Carlo samples are constant, $N=1000$, and we assume that the Monte Carlo likelihood approximation error is ignorable. Note that the estimated pdf for $\gamma = \pm 1$ is non-zero, but these boundary values are not ``unacceptable'' values, as discussed in \citet{Azzalini1999} and \citet{Azzalini2005}, as they represent a truncated Gaussian random field in the same way as $\gamma=0$ represents a Gaussian random field. The figure also displays that the occurrence of $\hat \gamma = \pm 1$ decreases with higher $p$, as discussed in \citet{Liseo1990}. The maximum likelihood estimates in Figure \ref{fig:synthetic_multiple} are not unbiased, but the estimators appear as consistent since the biases and variances shrink with increasing size of 
the random field $p$.

\section{Seismic inversion of data from the North Sea}

\begin{figure}%
\centering
\includegraphics[width=0.8\textwidth]{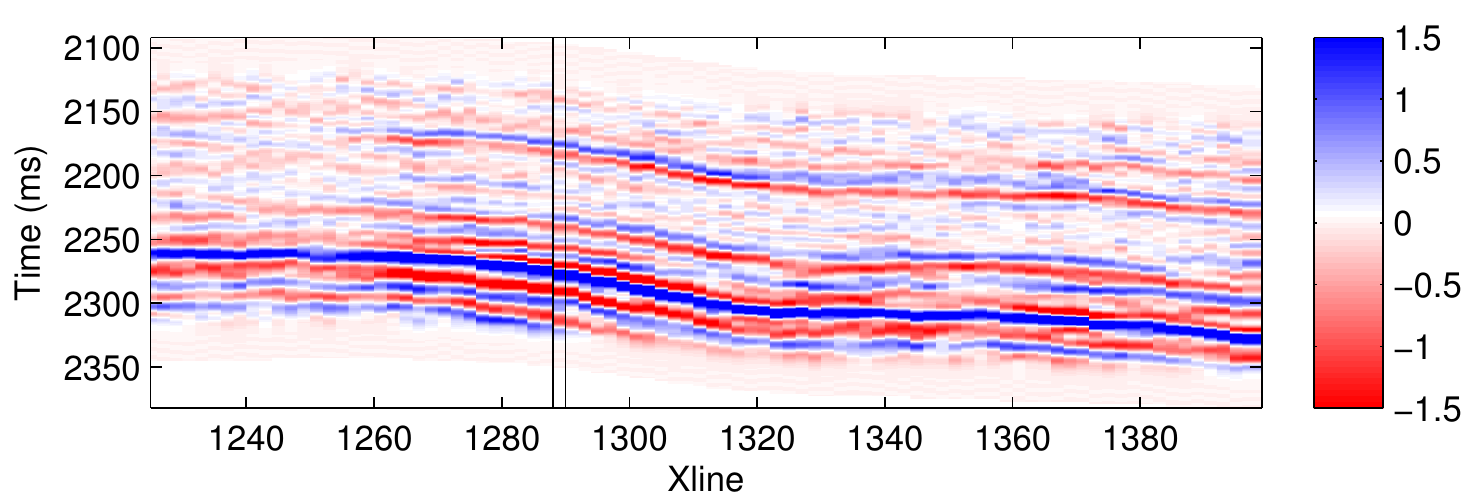}
\caption{Seismic amplitude data for angle $31 \ensuremath{^\circ}$ on a $125 \times 88$ grid. The depth is measured in seismic two-way travel-time. The well location is marked at around trace 1290.}%
\label{fig:seismic}%
\end{figure}

In this section we use the CSN random field model in a predictive setting. We consider inversion of seismic amplitude-versus-offset (AVO) data into elastic material properties (pressure-wave velocity, shear-wave velocity, density) in the subsurface, which is a major challenge in modeling of hydrocarbon reservoirs. The seismic AVO data are measurements from the Sleipner \O st field in the North Sea, which is a gas condensate field in the southern part of the North Sea. The depth of the reservoir is in the range from 2270 to 2500 meter sub-sea. We have seismic AVO data from a 2D profile and observations of the elastic material properties from one well, drilled through the reservoir, see Figure \ref{fig:seismic}.

In \citet{buland:185} the seismic inversion is casted in a Bayesian predictive setting with a Gaussian prior on the logarithm of the elastic material properties and Gauss-linear likelihood for the seismic observations. The methodology is illustrated on data from the Sleipner \O st field. To justify the use of a Gauss-linear model the prior model for the logarithm of the elastic material properties have to be assumed Gaussian, but these assumptions do not fit the observations from an available well particularly good. Data from the same area is also considered in \citet{karimi:R1} where a CSN model with a pseudo-likelihood is used on a 1D profile along one well. The pseudo-likelihood approach will suffer from instabilities in the parameter inference, especially in a 2D setting where the spatial coupling is more prominent due to strong horizontal correlation. In this section we consider the full 2D profile of the seismic data and aim at predicting the associated elastic material properties. We use the CSN 
random field presented above as prior model for the 2D profile of the logarithm of the elastic material properties. 

The seismic AVO data $\mathbf d$ are collected by a seismic survey, which is an active acoustic data acquisition technique. Explosions are fired at several locations at the surface and reflections from a grid covering the subsurface for a set of reflection angles are collected.  The data $\mathbf d$ represent angle-dependent seismic AVO data for three angles $[12 \ensuremath{^\circ}, 22 \ensuremath{^\circ}, 31 \ensuremath{^\circ}]$ at each grid node in a $(n_t \times n_x)$-grid covering the 2D profile. The dimension of the seismic AVO data is $ 3 \times n_t \times n_x = 3 \times 125 \times 88 = 33 \; 000 $. 

\begin{figure}%
\centering
\subfloat[][]{\includegraphics[width=0.7\textwidth]{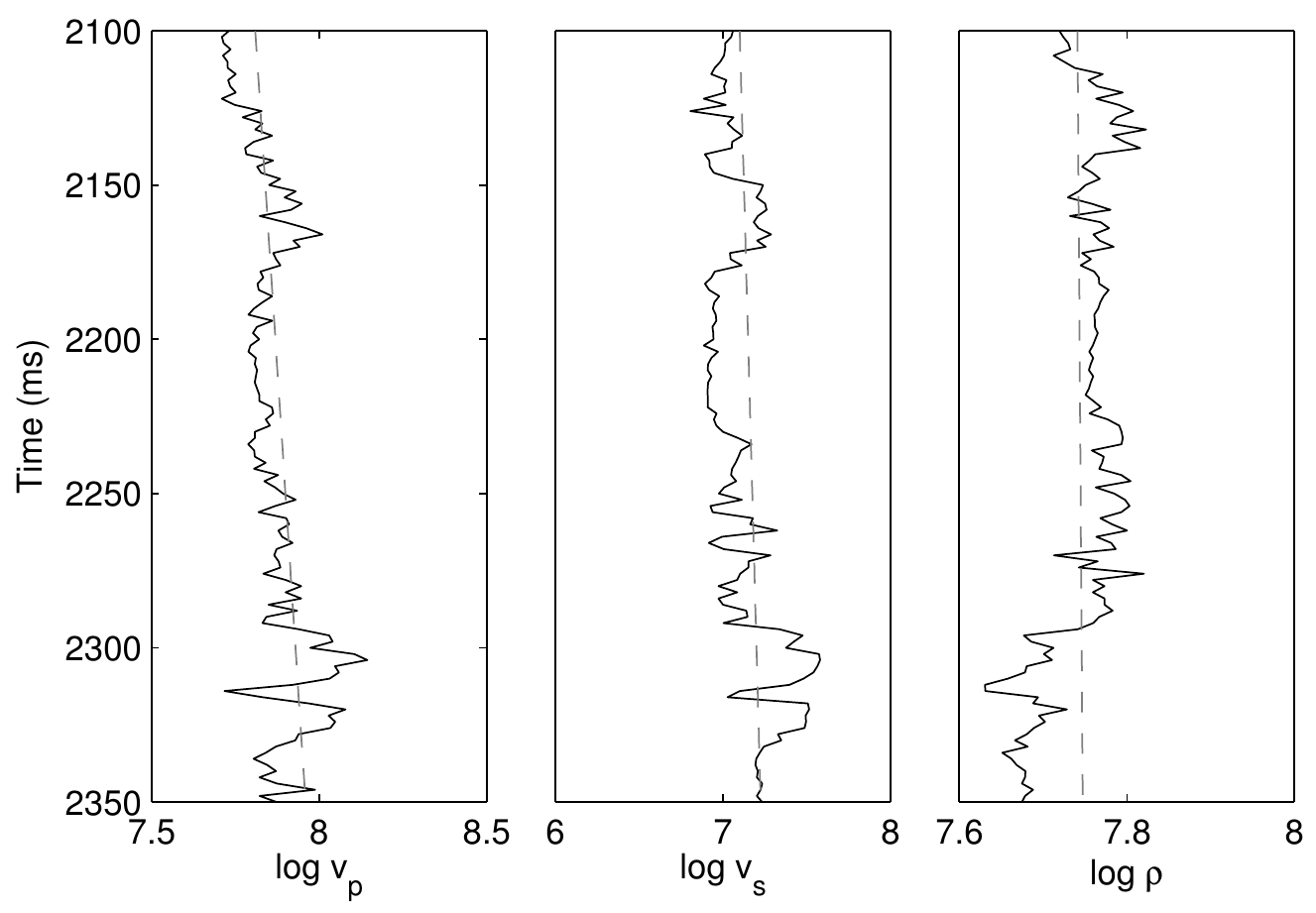}}%
\qquad
\subfloat[][]{\includegraphics[width=0.7\textwidth]{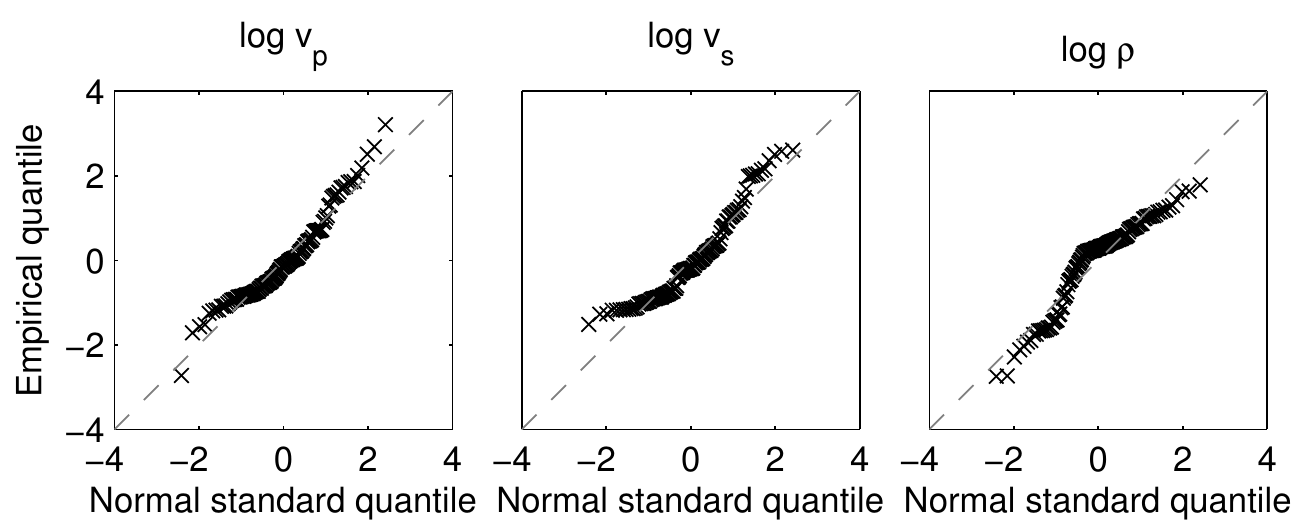}}%
\qquad
\subfloat[][]{\includegraphics[width=0.7\textwidth]{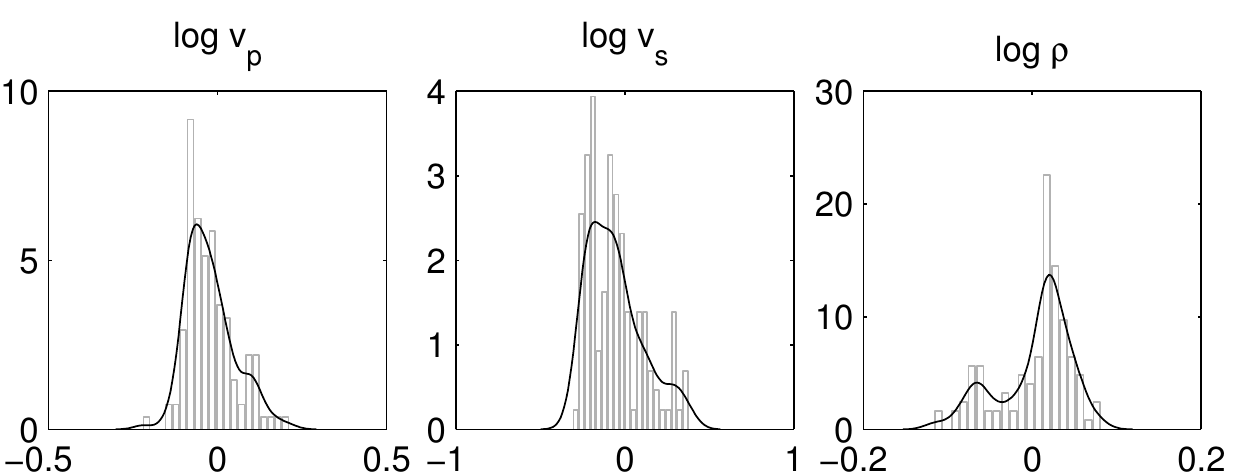}}%
\caption{Well observations of logarithm of pressure-wave $v_p$, share wave $v_s$, and density $\rho$. (a) Elastic properties in the well with a dashed estimated linear vertical trend, (b) quantile-quantile plot of empirical quantiles from data (residuals after linear trends removed) versus theoretical quantiles from normal distribution, (c) histogram and density plot of residuals.}%
\label{fig:welldata}%
\end{figure}

The variable of interest $\mathbf m$ represents the logarithm of the elastic material properties $(v_p,v_s,\rho)$ on the $(n_t \times n_x)$-grid covering the 2D profile. Hence, the dimension of $\mathbf m$ is $3 \times n_t \times n_x = 3 \times 125 \times 88 = 33 \; 000 $. The logarithmic transformation is used to get a linear relationship between the variables of interest $\mathbf m$ and the seismic data $\mathbf d$. The elastic material properties are also observed along the well trace, see Figure \ref{fig:seismic}. The observations in the well $\mathbf m_w$ are assumed to be exact and they are displayed in Figure \ref{fig:welldata}. The elastic material properties are centered around linear vertical trends, and their histograms are displayed in Figure \ref{fig:welldata}(b) and \ref{fig:welldata}(c). The histograms appear as skewed, or may even bimodal. 

The link between the observations and the elastic material properties, termed seismic likelihood model $[\mathbf d \mid \mathbf m]$, is defined by a weak-contrast, convolutional, linearized Zoeppritz model \citep{buland:185}. The convolutional forward model is defined by  the matrix $\mathbf G=\mathbf{WAD}$, where $\mathbf W$ is a convolutional matrix defined by the kernels, presented in Support S2, $\mathbf A$ is a matrix of angle-dependent weak contrast Aki-Richards coefficients \citep{akirickards}, and $\mathbf D$ is a differential matrix which calculates contrasts. This forward matrix $\mathbf G=\mathbf{WAD}$ represents the physics of the wave reflections. The reflection depends on the contrasts in the material properties ($\mathbf D$), it is angle dependent ($\mathbf A$), and wave propagation is a diffusive process ($\mathbf W$). The model is $\mathbf d = \mathbf G \mathbf m + \mathbf e$, where $\mathbf e$ is assumed to be a colored Gaussian error term with zero mean and covariance matrix $\boldsymbol 
\Sigma_{\mathbf e}$. The covariance matrix is parameterized as $\boldsymbol \Sigma_{\mathbf e} = \sigma^2_{\mathbf e} \cdot \mathbf C^w_{\mathbf e} \otimes \mathbf C^h_{\mathbf e} \otimes \mathbf C^v_{\mathbf e}$, where $\otimes$ denotes the Kronecker product, $\sigma^2_{\mathbf e}$ is the 
error variance, $\mathbf C^w_{\mathbf e}$ is a wavelet correlation matrix, $\mathbf C^h_{\mathbf e}$ is a horizontal correlation matrix, and $\mathbf C^v_{\mathbf e}$ is a vertical correlation matrix. These matrices are parameterized by exponential correlation matrices with parameters $d^w_{\mathbf e}$, $d^h_{\mathbf e}$, $d^v_{\mathbf e}$, respectively. Thus the likelihood is $p(\mathbf d \mid \mathbf m) = N(\mathbf G \mathbf m, \boldsymbol \Sigma_{\mathbf e})$.

The objective is to predict the elastic variables $\mathbf m$, from the observed seismic AVO data, $\mathbf d$. The variables of interest, $\mathbf m$, is high-dimensional, and so is the observations, $\mathbf d$. The forward matrix $\mathbf G$ is ill-conditioned and rank-deficient, however. Moreover, there is a colored error term, $\mathbf e$, in the observations. Hence $\mathbf m$ cannot be uniquely determined by $\mathbf d$. We cast this prediction in a Bayesian setting; hence the posterior distribution is the objective of the study
\begin{align}
p(\mathbf m \mid \mathbf d) = \mbox{const} \times p(\mathbf d \mid \mathbf m) \; p(\mathbf m), \label{eqn:seismic_posterior}
\end{align}
where $\mbox{const}$ is a normalizing constant and $p(\mathbf m)$ is the prior distribution of $\mathbf m$ which must be defined.

Let the prior model for $\mathbf m$ be a stationary CSN random field with $q = p = 3 \times n_t \times n_x = 3 \times 125 \times 88 = 33 \; 000 $ as previously defined: $p(\mathbf m) = CSN_{p,p}(\boldsymbol \mu_{\mathbf m}, \boldsymbol \Sigma_{\mathbf m}, \boldsymbol \Gamma,\mathbf 0, \boldsymbol \Delta)$, where $\boldsymbol \mu_\mathbf m = \boldsymbol \mu^0_\mathbf m \otimes \mathbf 1$, $ \boldsymbol \mu^0_\mathbf m = [\mu_{v_p}, \mu_{v_s}, \mu_\rho]^T$, and $\mathbf 1 \in \mathbb{R}^{n_x n_t \times 1}$. The covariance matrix $\boldsymbol \Sigma_{\mathbf m}$ is parameterized as $\boldsymbol \Sigma_{\mathbf m} = \boldsymbol \Sigma^0_{\mathbf m} \otimes \mathbf C_{\mathbf m}$, where $\boldsymbol \Sigma^0_{\mathbf m} \in \mathbb{R}^{3 \times 3}$ is the inter-variable covariance matrix and $\mathbf C_{\mathbf m} = \mathbf C_{\mathbf m}^h \otimes \mathbf C_{\mathbf m}^v $, $\mathbf C_{\mathbf m}^h \in \mathbb{R}^{n_x \times n_x}$ is a horizontal direction exponential correlation matrix with parameter 
$d^h_\mathbf m$ and $\mathbf C_{\mathbf m}^v \in \mathbb{R}^{n_t \times n_t}$ is an vertical direction exponential correlation matrix with parameter $d^v_\mathbf m$. The skewness parameter is $\boldsymbol \Gamma = \boldsymbol \Gamma^0 \otimes \mathbf I_{n_t n_x}$, where $\boldsymbol \Gamma^0 = \mbox{diag}(\boldsymbol \gamma), \; \boldsymbol \gamma =  [\gamma_{v_p}
, \gamma_{v_s}, \gamma_{\rho}]^T$, and $\mathbf \Delta = (\mathbf I_3-\mathbf \Gamma^0)  (\mathbf I_3-\mathbf \Gamma^0) \otimes \mathbf I_{n_t n_x}$.  The full $(2p \times 2p)$ covariance matrix for $p(\mathbf m)$ is
\begin{align}
\left[
\begin{array}{cc}
\boldsymbol \Sigma^0_{\mathbf m} \otimes \mathbf C_{\mathbf m} 		
& -\left(\boldsymbol \Sigma^0_{\mathbf m}  \left( \boldsymbol \Gamma^0 \boldsymbol \Omega^0_\mathbf m \right)^T \right) \otimes \mathbf C_{\mathbf m} \\
-\left(\boldsymbol \Gamma^0 \boldsymbol \Omega^0_\mathbf m \boldsymbol \Sigma^0_{\mathbf m}\right)  \otimes \mathbf C_{\mathbf m}	& \mathbf \Delta + \left(\left(\boldsymbol \Gamma^0 \boldsymbol \Omega^0_\mathbf m \right) \boldsymbol \Sigma^0_{\mathbf m} \left( \boldsymbol \Gamma^0 \boldsymbol \Omega^0_\mathbf m \right)^T\right) \otimes \mathbf C_{\mathbf m}
\end{array}
\right],
\end{align}
where $\boldsymbol \Omega^0_\mathbf m$ is a diagonal matrix where the elements are the square root of the inverse of the diagonal matrix of $\boldsymbol \Sigma^0_{\mathbf m}$, and is used to scale the covariance matrix of the truncated field, i.e. $\boldsymbol \Omega^0_\mathbf m \boldsymbol \Sigma^0_{\mathbf m} \left( \boldsymbol \Omega^0_\mathbf m \right)^T$ is a correlation matrix. This expression corresponds to Expression \ref{eqn:covariance} extended to a tri-variate random field.

% \begin{figure}
% 
% \begin{center}
%  \includegraphics{./fig/figure_8.eps}
% \end{center}
%  
% \caption{Graphical illustration of elastic parameters $\mathbf m$, seismic observations $\mathbf d$, and truncation parameter $\mathbf u$. }
% \label{fig:DAG}%
% \end{figure} 
% 
% The model can be graphically displayed by the graph in Figure \ref{fig:DAG}, where we have the elastic parameter $\mathbf m$, the seismic observations $\mathbf d$, and truncation parameters $\mathbf u$. Thus, the predictive distribution of the CSN random field is $p(\mathbf m \mid \mathbf d, \mathbf u < \mathbf 0)$ and the predictive distribution of the Gaussian random field is $p(\mathbf m \mid \mathbf d)$, where of course the parameter values for $\mathbf m$ are different for the CSN and Gaussian model.

The unknown model parameters in both the likelihood and the prior models are $\boldsymbol \theta = $ $(\sigma^2_{\mathbf e}$, $d^w_{\mathbf e}$, $d^h_{\mathbf e}$, $d^v_{\mathbf e}$, $\boldsymbol \mu^0_\mathbf m$, $\boldsymbol \Sigma^0_{\mathbf m}$, $\boldsymbol \gamma$, $d^h_\mathbf m$, $d^v_\mathbf m )$. We estimate $\boldsymbol \theta $ by maximum marginal likelihood for seismic AVO data and well observations $(\mathbf d, \mathbf m_w)$. These parameter estimates, $\hat{\boldsymbol \theta}$, are used as plug-in values in the posterior model $p(\mathbf m \mid \mathbf d)$, see Expression \ref{eqn:seismic_posterior}, to obtain an operable model. 

The marginal likelihood to be maximized with respect to $\boldsymbol \theta$ is: 
\begin{align}
p(\mathbf d, \mathbf m_w ; \boldsymbol \theta ) 
= \int p(\mathbf d, \mathbf m ; \boldsymbol \theta)  \; \mathrm d \mathbf m_{-w} 
= \int p(\mathbf d ; \mathbf m , \boldsymbol \theta) \; p(\mathbf m ; \boldsymbol \theta)  \; \mathrm d \mathbf m_{-w}, \label{eqn:marginal_likelihood}
\end{align}
where $\mathbf m_{-w}$  denotes the material properties everywhere except in the well trace. Expression \ref{eqn:marginal_likelihood} is analytically tractable since CSN random fields are closed under linear operations.
In practice, we only use 20 seismic traces on each side of the well to reduce the computational burden, i.e. the normalizing constant integral in the CSN distributions has dimension $q = 125 \times 41 \times 3 = 15\; 375$. 

The estimation procedure is identical to the one discussed in the parameter estimation section, and we use $N = 10\; 000$ Monte Carlo samples. Note that we use a common grid design for both predictions and model parameter inference in order to obtain consistent estimates. Each likelihood evaluation takes a couple of minutes on the laptop computer; thus the optimization procedure takes a couple of hours. The estimated parameters for the CSN random field prior model are
\begin{align*}
&\sigma^2_{\mathbf e} = 0.27,  \quad d^w_{\mathbf e} = 0.09,  \quad d^h_{\mathbf e} = 22.85,  \quad d^v_{\mathbf e} = 16.19, \\
&\boldsymbol \mu^0_\mathbf m 
=
\left[
\begin{array}{r}
   -0.27 \\
   -0.55 \\
    0.10
\end{array}
\right],
\quad
\boldsymbol \Sigma^0_m 
=
\left[
\begin{array}{rrr}
    0.0052  &   0.0080  &   0.0003 \\
    0.0080  &   0.0342  &  -0.0010 \\
    0.0003  &  -0.0010  &   0.0022 \\
\end{array}
\right],
\quad
\boldsymbol \gamma 
= 
\left[
\begin{array}{r}
 0.941 \\
 0.996 \\
-0.902
\end{array}
\right], \\
&d^h_\mathbf m = 3.37, \quad d^v_\mathbf m = 13.76. 
\end{align*}
Note that the skewness parameter $\boldsymbol \gamma $ indicates a positive skewness for pressure-wave velocity $v_p$ and shear-wave velocity $v_s$, and negative skewness for density $\rho$, which agree with the histograms in Figure \ref{fig:welldata}. Note also that the correlation in the horizontal direction $d^h_\mathbf m$ of the prior model for the elastic properties $\mathbf m$ is much higher than in the vertical direction $d^v_\mathbf m$.

We will compare the CSN random field model with a Gaussian random field model, which is known to be a CSN random field with $\boldsymbol \gamma = \mathbf 0$. The estimated parameters for the Gaussian model are
\begin{align*}
&\sigma^2_{\mathbf e} = 0.32,  \quad d^w_{\mathbf e} = 0.10,  \quad d^h_{\mathbf e} = 24.75,  \quad d^v_{\mathbf e} = 17.9, \\
& \boldsymbol \mu^0_\mathbf m 
=
\left[
\begin{array}{r}
   -0.025 \\
   -0.054 \\
    0.0010
\end{array}
\right],
\quad
\boldsymbol \Sigma^0_m 
=
\left[
\begin{array}{rrr}
    0.0040  &  0.0035 &   0.0002 \\
    0.0035  &  0.0193 &  -0.0009 \\
    0.0002  & -0.0009 &   0.0013 \\
\end{array}
\right], \\
&d^h_\mathbf m = 3.25, \quad d^v_\mathbf m = 14.85.
\end{align*}
The estimation procedure for the Gaussian model takes only a couple of minutes on the laptop computer. 
The main deviations between the estimates under the two models are the differences in the location $\boldsymbol \mu_\mathbf m^0$ and scale $\boldsymbol \Sigma_\mathbf m^0$ parameter estimates. Recall that the location $\boldsymbol \mu_\mathbf m^0$ and scale $\boldsymbol \Sigma_\mathbf m^0$ parameters are not identical to the expected value and variance in the CSN model.

Having an estimate of the model parameters, $\hat {\boldsymbol \theta}$, we use $p(\mathbf m \mid \mathbf d)$ with plug-in values $\boldsymbol{ \hat \theta}$ as the predictive posterior distribution, see Expression \ref{eqn:seismic_posterior}. The posterior model is a CSN random field and analytically tractable due to the closure properties of the CSN distribution \citep{Gonzalez-Farias2004,karimi:R1}. Note that we do not use the observations of the elastic material properties in the well trace explicitly when predicting $\mathbf m$, the well observations $\mathbf m_w$ will only have influence on the predictions through $\boldsymbol{ \hat \theta}$.

\begin{figure}%
\centering
\subfloat[][]{\includegraphics[width=0.7\textwidth]{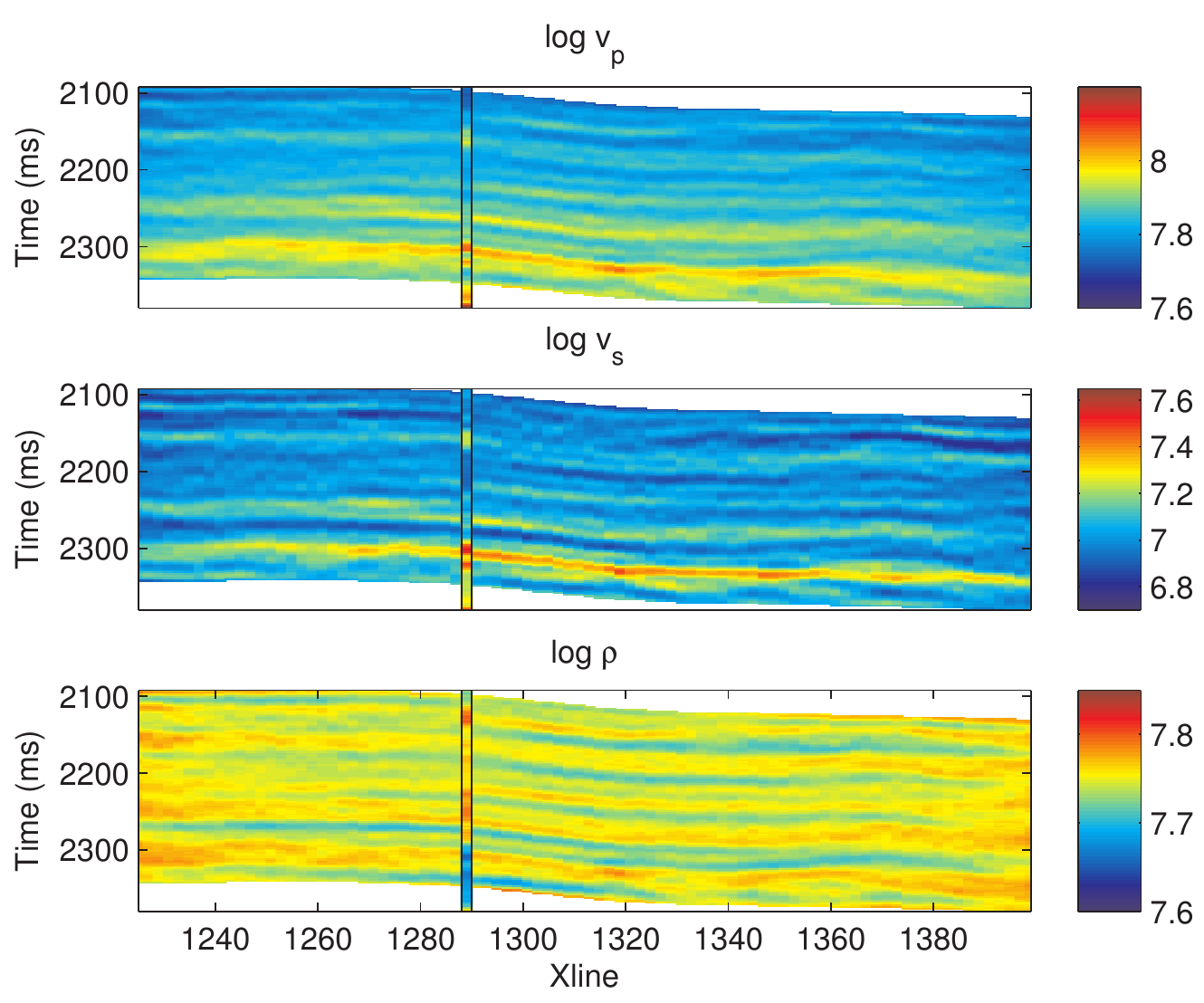}}%
\qquad
\subfloat[][]{\includegraphics[width=0.7\textwidth]{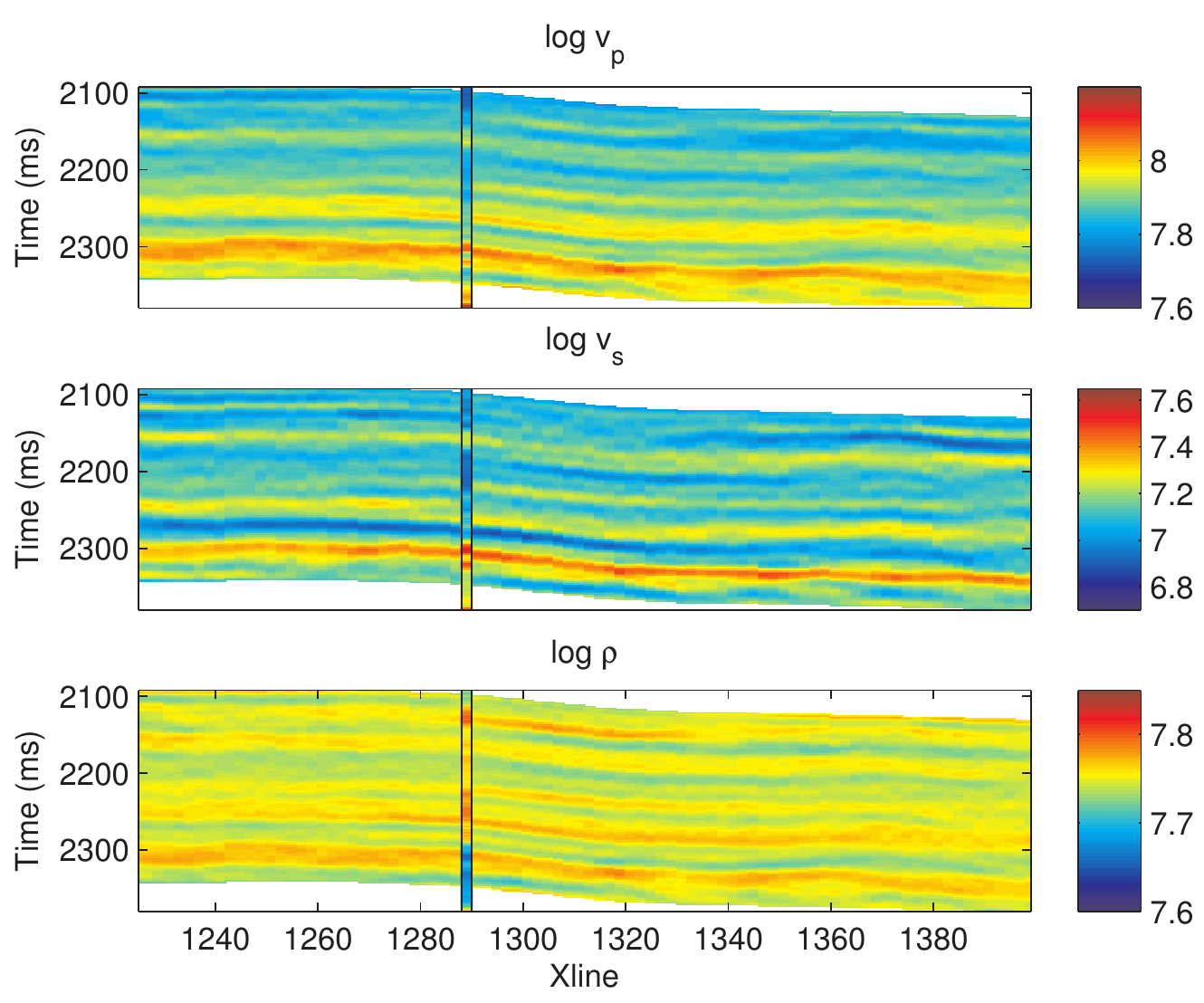}}%
\caption{Posterior median, the well observations are marked in well trace. (a) CSN random field model, (b) Gaussian random field model.}%
\label{fig:mean}%
\end{figure}

The posterior distribution is estimated by sampling $10\; 000$ samples using the MH algorithm in Appendix A. Figure \ref{fig:mean} displays the posterior median for both models. The predictions for the two models appear as fairly similar, but the predictions with the CSN model have generally lower values for $v_p$ and $v_s$ compared to the Gaussian model. The predictions for $\rho$ appears to deviate more from the well observations for both models than the predictions for $v_p$ and $v_s$, but this is expected from the geophysical model since there is less information about $\rho$ in the data. Realizations from the CSN and Gaussian posterior distributions are presented in Support S2, and the well observations do not deviate dramatically.

The posterior standard deviations for the CSN model are both observation design and value dependent, see Support S2. Usually higher prediction variance for extreme predictions. The posterior standard deviations for the Gaussian model are only dependent on the observations design, not the observed values, hence they are almost constant due to a symmetric design. The standard deviations for the Gaussian model is slightly larger than the typical values in the CSN model for $v_p$ and $v_s$, while standard deviations for $\rho$ are similar in both models.

\begin{figure}%
\centering
\subfloat[][]{\includegraphics[width=0.47\textwidth]{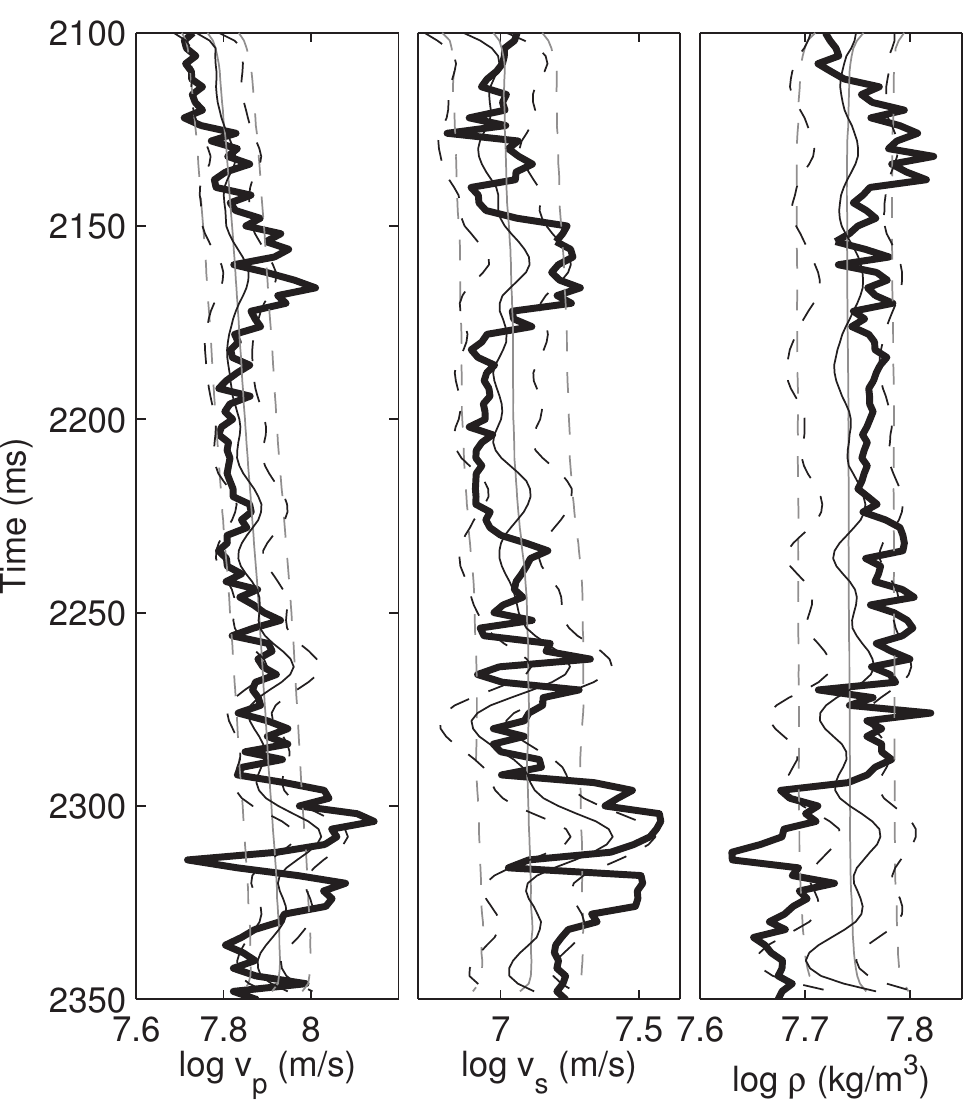}}%
\qquad
\subfloat[][]{\includegraphics[width=0.47\textwidth]{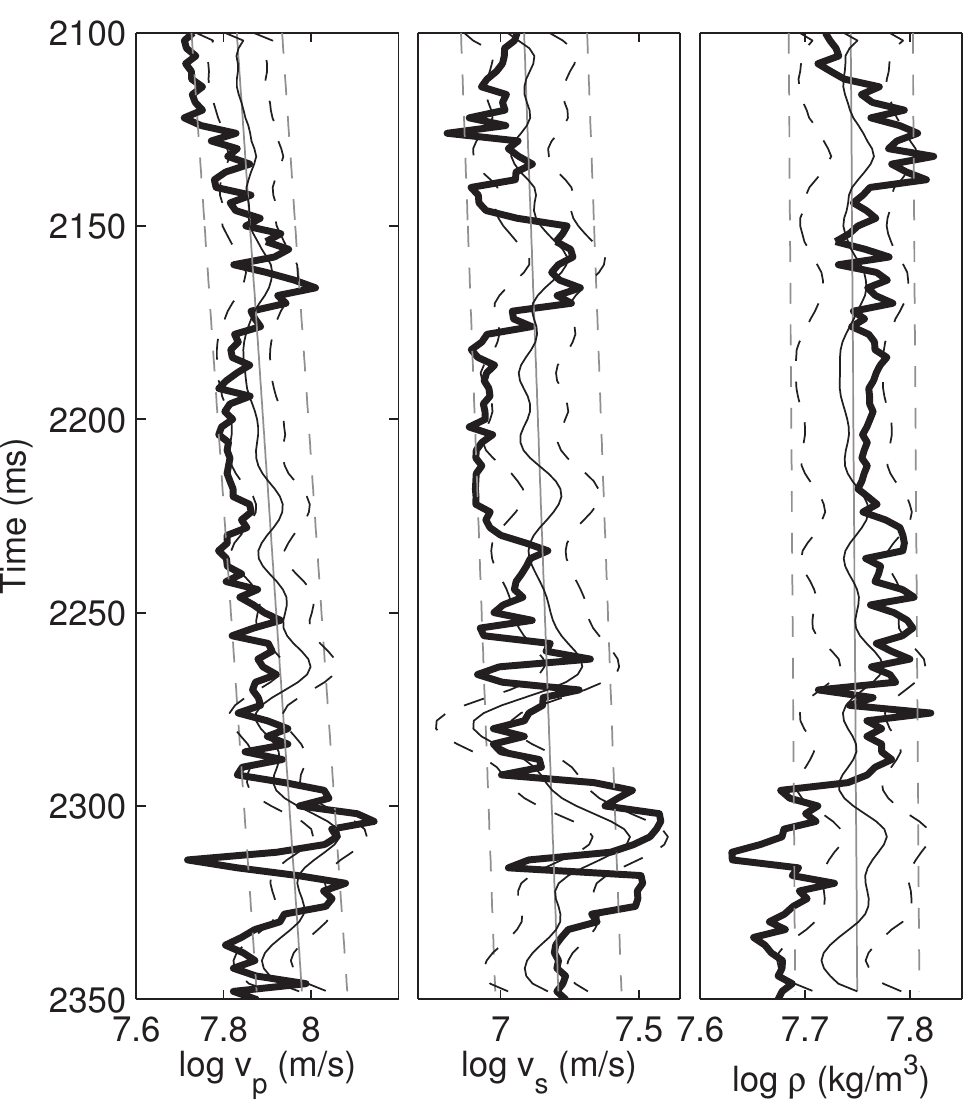}}%
\caption{Median predictions of elastic material properties in well trace. Bold black solid curve is well observations, thin black solid curve is posterior median, dashed black line is posterior 80\% prediction interval, thin gray solid is prior median, dashed gray is prior 80\% prediction interval. (a) CSN  random field model, (b) Gaussian random field model.}%
\label{fig:well}%
\end{figure}

Figure \ref{fig:well} displays the observations $\mathbf m_w$ and the well predictions from the seismic data. The bold black solid lines are well observations, the thin black solid lines are posterior medians, dashed black lines are posterior 80\% prediction intervals, the thin gray solid lines are prior medians, and the dashed gray lines are prior 80\% prediction intervals. We see that the CSN model predictions match the well observations better than the Gaussian model for low values of $v_p$ and $v_s$, and reach almost as high as the Gaussian model predictions for the high values. This is as expected since the CSN model has larger flexibility than the Gaussian model. The differences for $\rho$ predictions are small. The CSN model produce asymmetric predictions intervals due to skewness in the marginal posterior pdfs, but this effect is not very prominent in the display.

\begin{table}
\begin{center}
\begin{tabular}{|c|rr|rr|rr|}
\hline
& \multicolumn{2}{|c|}{MAE} & \multicolumn{2}{|c|}{Prior coverage} & \multicolumn{2}{|c|}{Posterior coverage} \\
& CSN & Gaussian & CSN & Gaussian & CSN & Gaussian \\
\hline
$v_p$	&	0.044	& 0.065 & 0.74  & 0.78	&	 0.68 & 0.43  \\
$v_s$	&	0.105 	& 0.133 & 0.80  & 0.81	&	 0.66 & 0.58  \\
$\rho$	&	0.028	& 0.034 & 0.73  & 0.81	&	 0.68 & 0.70  \\
\hline
\end{tabular}
\end{center}
\caption{Cross validation of median predictions and prediction intervals with respect to observations. Mean absolute error (MAE), posterior and prior coverage for the CSN random field and Gaussian random field model.}
\label{tbl:well}
\end{table}

Table \ref{tbl:well} displays the mean absolute error (MAE), prior and posterior coverage for the prediction intervals for the CSN and Gaussian models, with the well observations $\mathbf m_w$ used as truth. The MAE is reduced by $15-30\%$ when using the CSN model compared to the Gaussian model. The prior $80\%$ coverage is a reference coverage, and the posterior coverage shall ideally be identical to the prior one. The reduction in coverages, entailing underestimation of the prediction intervals, are much larger for the Gaussian model than for the CSN model for $v_p$ and $v_s$. Recall that $v_p$ and $v_s$ are the variables with most skewness. This indicates that the CSN model is superior to the Gaussian model in seismic inversion into elastic material properties for the Sleipner case.

\begin{figure}%
\centering
  \includegraphics[width=0.7\textwidth]{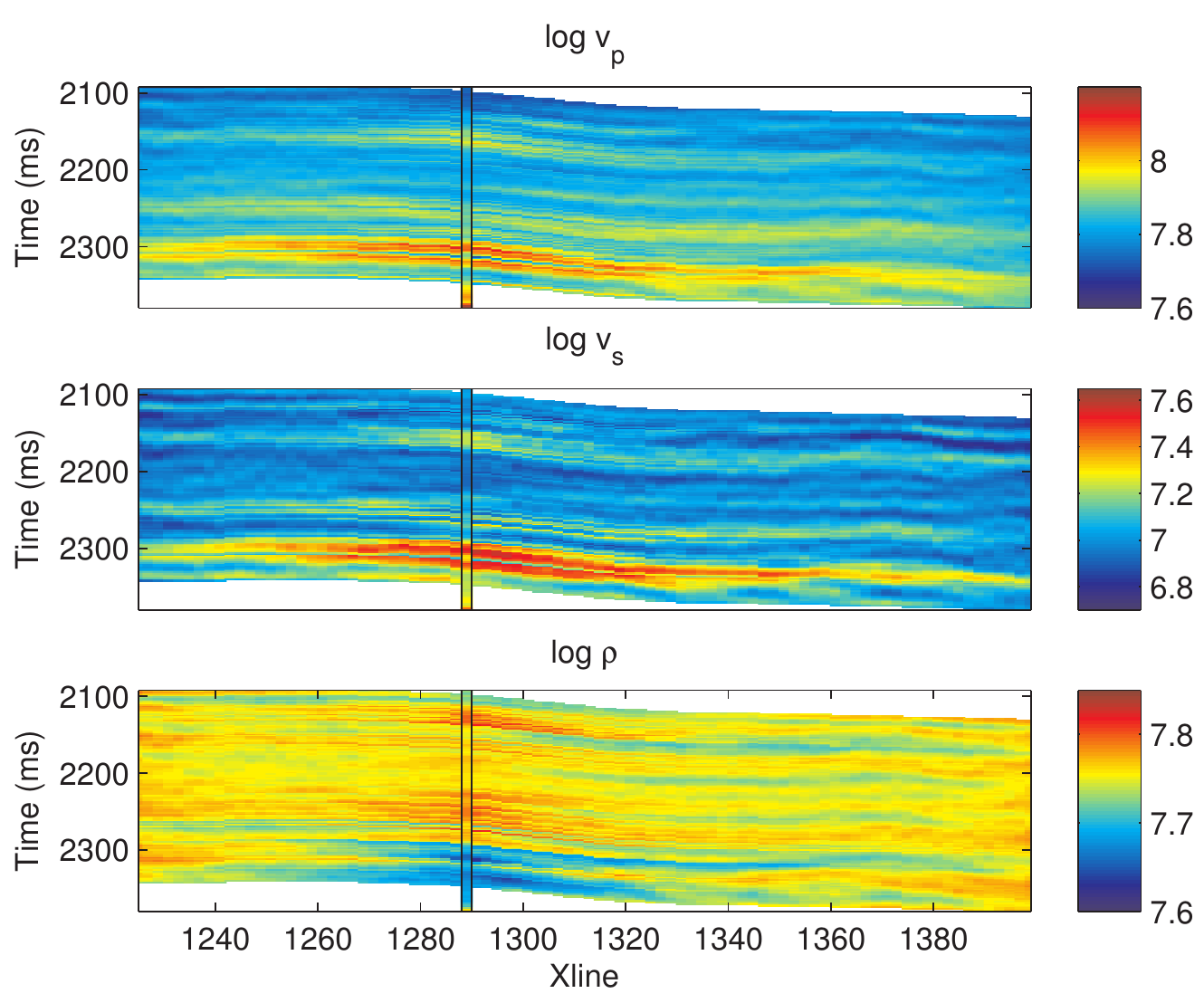}%
\caption{Posterior median conditioned on both seismic AVO data and the well observations for the CSN  random field model.}%
\label{fig:cond_mean}%
\end{figure}

The well observations $\mathbf m_w$ can also be used in the prediction based on the CSN random field model, hence the predictive distribution is  $p(\mathbf m \mid \mathbf d, \mathbf m_w )$. Figure \ref{fig:cond_mean} corresponds to Figure \ref{fig:mean}a when also $\mathbf m_w$ is conditioned on. As expected these predictions reproduce the well observations and appear with  higher resolution close to the well. A simulated realization from the posterior distributions is displayed in Support S2, and the well observations appear as an integral part of the realization.

For the Gaussian random field model, the computation time for parameter estimation and sampling are within minutes on the laptop computer, while for the CSN random field model the corresponding time is hours. One may ask whether the rather small improvements in the predictions are worth the large increase in computation time, although in some cases these minor improvements may be important for identifying a hydrocarbon reservoir of immense value.

\section{Conclusion}

We define an approximately stationary CSN random field with skewed Gaussian marginal distributions. The field is based on the definition of CSN multivariate distribution \citep{Gonzalez-Farias2004} and has a somewhat different parameterization than the CSN random field defined in \citet{Allard2007}. We demonstrate that there is a strong dependence between the maximum skewness in the marginal pdfs and the spatial coupling.

A Metropolis-Hastings algorithm for effective simulation of realizations from the CSN random field, and an procedure for estimating model parameters by maximum likelihood are presented. A simulation study on different CSN random fields illustrate that the maximum skewness of the marginal distributions is severely reduced compared to the skewness in univariate skew-normal distribution due to spatial coupling effects. The maximum likelihood estimates for the model parameters are biased for small random fields, but the bias and variances of the estimates are reduced with increasing extent of the random field. The model parameter estimators appear to be consistent.

A case study of seismic AVO inversion into elastic material properties is presented. The inversion is cast in a Bayesian predictive setting with a tri-variate CSN random field prior model and a Gauss-linear likelihood model. The posterior model is also a CSN random field which is analytically tractable. Plug-in model parameter estimates based on maximum marginal likelihood is used. $15-30\%$ improvements in MAE in  the predictions compared to a Gaussian random field model are documented.

The computational cost of parameter estimation and prediction is feasible even for random field discretized into grid of size at least $10^5$. Our example runs within hours on a laptop computer.

\section{Acknowledgments}
The research is a part of the Uncertainty in Reservoir Evaluation (URE) activity at the Norwegian University of Science and Technology (NTNU). We thank the operator Statoil and the Sleipner licence (Statoil, ExxonMobil, and Total) for providing the data.

\section*{Supporting Information}

Additional supporting information may be found in the online version of this article: \\
% S1. Algorithm: Sampling from a truncated multivariate normal distribution.\\
S1. Illustration: Marginals of truncated multivariate normal distributions.\\
% S3. Algorithm: Monte Carlo estimation of normal orthant probabilities.\\
S2. Figures: Additional figures from seismic inversion of data from the North Sea.

\section*{References}
\bibliographystyle{elsarticle-harv}
\bibliography{../share/ref}

% \vspace{2cm} 

% \noindent \textbf{Address author:} \\
% Kjartan Rimstad \\
% Department of Mathematical Sciences\\
% Norwegian University of Science and Technology\\
% Alfred Getz vei 1 \\
% 7491 Trondheim \\
% Norway \\
% rimstad@math.ntnu.no\vspace{0.7cm} \\

\appendix
\renewcommand*{\thesection}{\Alph{section}}

\section{Algorithm: Sampling from a truncated multivariate normal distribution} \label{sec:app_sample}

Consider the problem of sampling realizations from a truncated multivariate normal distribution with unnormalized density $I(\mathbf x \leq \mathbf 0) \times \phi_n(\mathbf x ; \boldsymbol \mu, \boldsymbol \Sigma)$, where $\mathbf x, \boldsymbol \mu \in \mathbb R^n$, $ \boldsymbol \Sigma \in \mathbb R^{n \times n}$, $I(\cdot)$ is the indicator function, the notation $\mathbf x \leq 0$ corresponds to all elements of $\mathbf x$ being jointly negative, and $\phi_n(\mathbf x ; \boldsymbol \mu, \boldsymbol \Sigma)$ is the $n$-dimensional multivariate normal density distribution. In order to sample from this distribution we extend the Metropolis-Hastings algorithm in \citet{Robert:TruncGaussian} with a block independent proposal distribution defined by
\begin{align}
p^*(\mathbf x^a \mid \mathbf x^b)
&= \prod_{i=1}^q I( x_i^a \leq 0) \; \frac{\phi_1(x_i^a \mid \mathbf x_{1:i-1}^a, \mathbf x^b ; \boldsymbol \mu, \boldsymbol \Sigma)}{\Phi_1(0 \mid \mathbf x_{1:i-1}^a,\mathbf x^b ; \boldsymbol \mu, \boldsymbol \Sigma)}
, \label{eqn:mcmcproposal}
\end{align}
where $n_a$ is the block size, $\mathbf x^a \in \mathbb{R}^{n_a}, \mathbf x^b \in \mathbb{R}^{n-n_a}$, $\phi_1(x_i^a \mid \mathbf x_{1:i-1}^a, \mathbf x^b ; \boldsymbol \mu, \boldsymbol \Sigma)$ and $\Phi_1( 0 \mid \mathbf x_{1:i-1}, \mathbf x^b  ; \boldsymbol \mu, \boldsymbol \Sigma)$ are the conditional normal probability and cumulative probability distribution of $x_i$ given $\mathbf x_{1:i-1}^a$ and $\mathbf x^b$, respectively, with $\mathbf x_{1:i-1} = (x_1,x_2, \ldots x_{i-1})$. Note that when $n_a=1$ we have the algorithm in \citet{Robert:TruncGaussian}. Note also that the distribution in Expression \ref{eqn:mcmcproposal} is normalized, it is easy to sample sequentially from this distribution, and that the distribution depends on the ordering of $\mathbf x$. Expression \ref{eqn:mcmcproposal} is the distribution which is used as independent sampler in \citet{Genz1992} for estimating orthant probabilities.

The acceptance probability in the accept/reject step is
\begin{align}
\alpha 
& = \min \left\lbrace 1, \frac{p({\mathbf x^a}' \mid \mathbf x^b)}{p(\mathbf x^a \mid \mathbf x^b)} \cdot \frac{p^*(\mathbf x^a \mid \mathbf x^b)}{p^*({\mathbf x^a}' \mid \mathbf x^b)} \right\rbrace \notag \\ 
& = \min \left\lbrace 1, \frac{ \prod_{i=1}^{n_a} \Phi_1( 0 \mid {\mathbf x^a_{1:i-1}}', \mathbf x^b  ; \boldsymbol \mu, \boldsymbol \Sigma)}{ \prod_{i=1}^{n_a} \Phi_1( 0 \mid \mathbf x^a_{1:i-1}, \mathbf x^b  ; \boldsymbol \mu, \boldsymbol \Sigma)} \right\rbrace,
\end{align}
where ${\mathbf x^a}'$ is the new proposed state.
The Metropolis-Hastings algorithm is presented in Algorithm \ref{alg1}.

\begin{algorithm}
\DontPrintSemicolon
\BlankLine\;
Initialize $\mathbf x\leq \mathbf 0$. \;
\textbf{Iterate} \;
\quad Choose one element $i$ at random in $\mathbf x$.  \;
\quad Find the set of the $n_a$ closest by correlation element to $i$. \; 
\quad Define the set of the $n_a$ elements $a_i$ and $b_i$ as it complement. \;
\quad Sample $\mathbf x'_{a_i \mid b_i} \sim p^*(\mathbf x^{a_i} \mid \mathbf x^{a_i})$. \;
\quad Accept $\mathbf x'_{a_i \mid b_i}$ with probability $\alpha$. \;
\textbf{End} \;
\BlankLine \;
\caption{Sampling from truncated multivariate normal distribution}
\label{alg1}
\end{algorithm}

In practice we calculate the conditional distributions in advance. To save memory and time we also limit the elements in $\mathbf x$, i.e. sets, we are allowed to choose, but we try to choose the allowed elements in a way such that all elements in $\mathbf x$ has approximately equal update probability. We normally use the block size $n_a = 100$.

\section{Algorithm: Monte Carlo estimation of normal orthant probabilities} 

Consider the problem of estimating the orthant probability 
\begin{align}
\Phi_n(\mathbf 0;\boldsymbol \mu, \boldsymbol \Sigma)
&= \int I(\mathbf x \leq \mathbf 0) \; \phi_n(\mathbf x;\boldsymbol \mu, \boldsymbol \Sigma) \; \mathrm d \mathbf x,
\end{align} 
where $\mathbf x, \boldsymbol \mu \in \mathbb R^n$, $ \boldsymbol \Sigma \in \mathbb R^{n \times n}$, $I(\cdot)$ is the indicator function, the notation $\mathbf x \leq 0$ corresponds to all elements of $\mathbf x$ being jointly less than or equal to zero,  and $\phi_n(\mathbf x ; \boldsymbol \mu, \boldsymbol \Sigma)$ is the $n$-dimensional multivariate normal density distribution.
The usual importance sampling Monte Carlo approximation with importance function $f_n(\mathbf x;\boldsymbol \mu, \boldsymbol \Sigma)$ is
\begin{align}
\Phi_n(\mathbf 0;\boldsymbol \mu, \boldsymbol \Sigma) 
&\approx \sum_{j=1}^N I(\mathbf x^j \leq \mathbf 0) \; \frac{\phi_n(\mathbf x^j;\boldsymbol \mu, \boldsymbol \Sigma)}{f_n(\mathbf x^j;\boldsymbol \mu, \boldsymbol \Sigma)},
\end{align}
with $\mathbf x^j \sim f_n(\mathbf x;\boldsymbol \mu, \boldsymbol \Sigma); \; j = 1,\ldots N$ and $N$ is the number of Monte Carlo sampling points. We follow the approach presented in \citet{Genz1992} and use the importance function
\begin{align}
f_n(\mathbf x;\boldsymbol \mu, \boldsymbol \Sigma)
&= \prod_{i=1}^n I( x_i \leq 0) \; \frac{\phi_1(x_i \mid \mathbf x_{1:i-1} ; \boldsymbol \mu, \boldsymbol \Sigma)}{\Phi_1(0 \mid \mathbf x_{1:i-1} ; \boldsymbol \mu, \boldsymbol \Sigma)}.
\end{align}
where $\phi_1(x_i \mid \mathbf x_{1:i-1}; \boldsymbol \mu, \boldsymbol \Sigma)$ and $\Phi_1( 0 \mid \mathbf x_{1:i-1}, \mathbf x  ; \boldsymbol \mu, \boldsymbol \Sigma)$ are the conditional normal probability and cumulative probability distribution of $x_i$ given $\mathbf x_{1:i-1}$, respectively, with $\mathbf x_{1:i-1} = (x_1,x_2, \ldots x_{i-1})$.
However, we also introduce a mean shift parameter $\boldsymbol \eta$ in the importance function. Then the importance sampling approximation appear as
\begin{align}
\Phi_n(\mathbf 0;\boldsymbol \mu, \boldsymbol \Sigma) 
& \approx \sum_{j=1}^N \frac{\phi_n(\mathbf x^j;\boldsymbol \mu, \boldsymbol \Sigma)}{\phi_n(\mathbf x^j;\boldsymbol \mu + \boldsymbol \eta , \boldsymbol \Sigma)} \prod_{i=1}^n \Phi_1(0 \mid \mathbf x^j_{1:i-1} ; \boldsymbol \mu + \boldsymbol \eta, \boldsymbol \Sigma), 
\end{align}
with $\mathbf x^j \sim f_n(\mathbf x;\boldsymbol \mu+ \boldsymbol \eta, \boldsymbol \Sigma), \; j = 1,\ldots N$. We use $\eta_i \approx -1.8 \Sigma_{ii}$ when the correlation structure in $\boldsymbol \Sigma$ is high, and close to $0$ when $\boldsymbol \Sigma$ is a diagonal matrix. It is also possible to use a different covariance matrix in the importance function, but the variance reduction we were able to attain was small compared to the extra computational time.



\section*{Supplementary material}

# Skew-Gaussian Random Fields

Kjartan Rimstad[a,], Henning Omre[a]

[a]*Department of Mathematical Sciences, Norwegian University of Science and Technology, Trondheim, Norway*

*Email addresses:* rimstad@gmail.com (Kjartan Rimstad), omre@math.ntnu.no (Henning Omre)

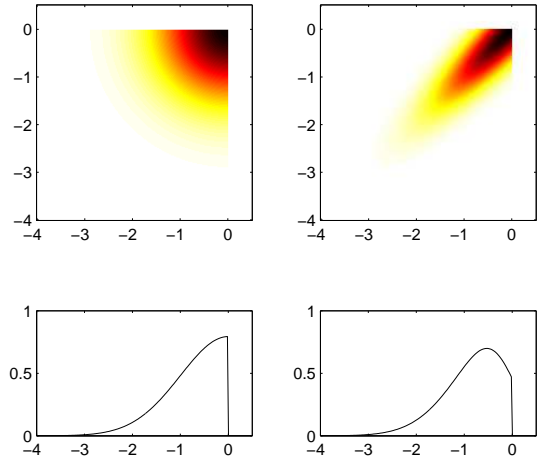

Figure 1: Interaction between trucation and correlation. First column: bivariate density and marginal of two independent truncated normal variables. Second column: bivariate density and marginal of two truncated normal variables with correlation 0.9.

## Support S1. Illustration: Marginals of truncated multivariate normal distributions

Consider the latent truncated Gaussian random field. In a univariate truncated Gaussian case, e.g. white noise Gaussian random field, the restriction is only that the associated univariate latent variable has to be negative. When we consider the multivariate truncated Gaussian random field, then spatial correlation is introduced, which in our case is positive. Thus, in addition to that the associated univariate latent variable has to be negative, also the other correlated latent variables have to be negative, which reduces the probability to have latent values close to the truncation border. This effect is illustrated in Figure 1 where the marginal distribution of two truncated normal distributed random variables with zero and 0.9 correlation is displayed. We see that the mode moves away from the truncation border. This effect increases when we introduce even more correlated variables, and the marginal distribution becomes more similar to the univariate normal distribution.

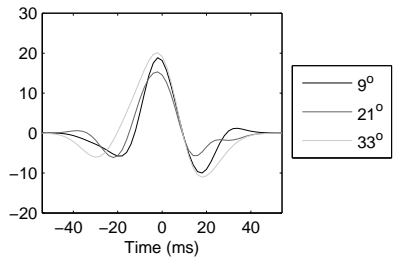

Figure 2: Vertical wavelet functions - supplied from the data-set owner.

# Support S2. Figures: Additional figures from seismic inversion of data from the North Sea

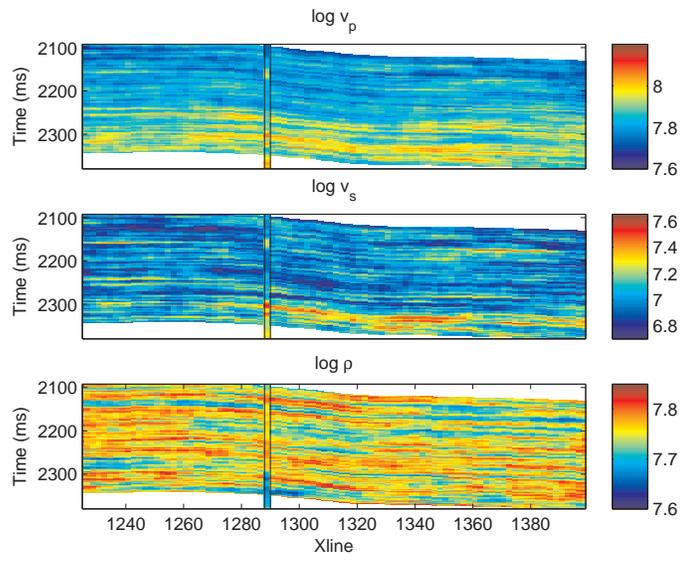

(a)

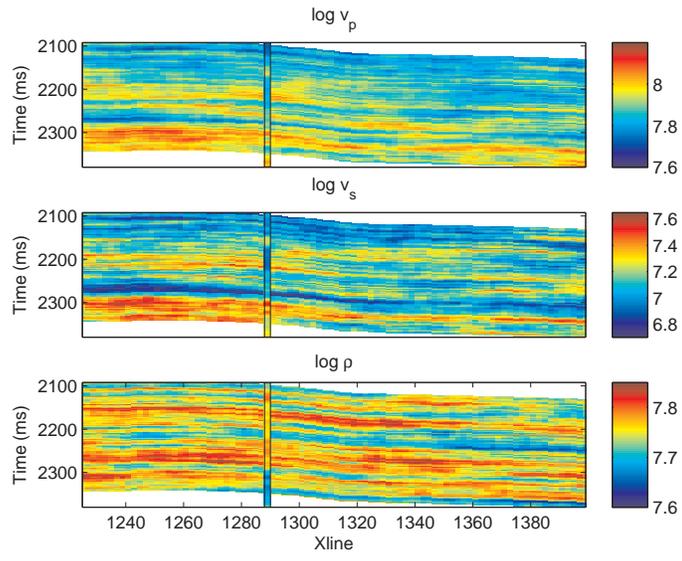

(b)

Figure 3: Simulated realizations from posterior distribution, the well observations are is marked in well trace. (a) CSN random field model, (b) Gaussian random field model.

4

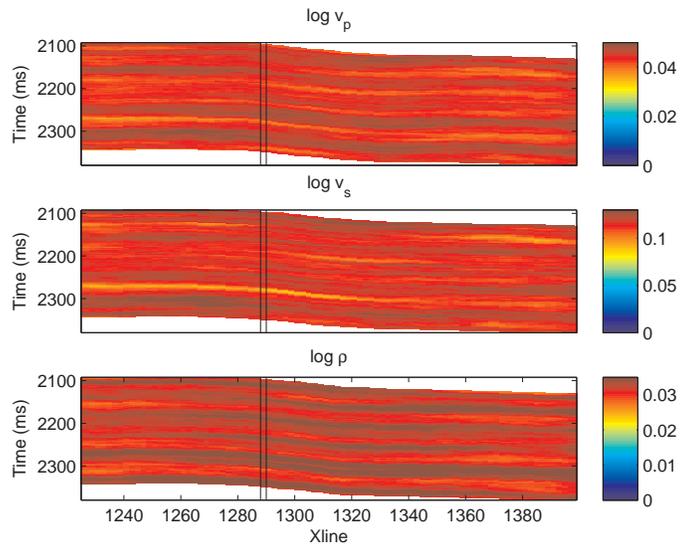

Figure 4: Standard deviation from the posterior distribution for CSN random field model. For the Gaussian random field model the standard deviations are approximately constant due to the symmetric observations design, they are $(130, 146, 71)$ for $(\log v_p, \log v_s, \log \rho)$, respectively

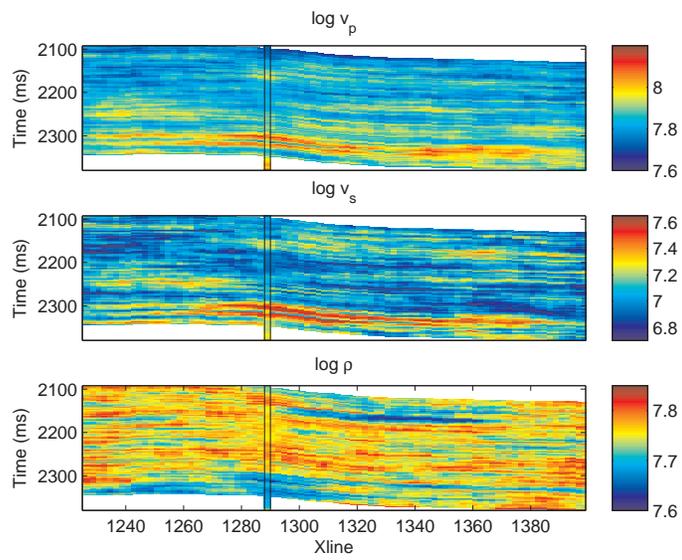

Figure 5: Simulated realization from the posterior distribution conditioned on both seismic AVO data and the well observations for the CSN random field model.



\end{document}